\documentclass[%
reprint,
aps,
pra,
letterpaper,
twocolumn,
superscriptaddress,
]{revtex4-2}

\usepackage[utf8]{inputenc}
\usepackage{amsmath,amsfonts,amssymb}
\usepackage{mathtools}
\usepackage{bm}
\usepackage[caption=false,justification=justified]{subfig}
\usepackage[colorlinks,unicode]{hyperref}
\usepackage{enumitem}
\usepackage{dsfont}
\usepackage[dvipsnames]{xcolor}
\usepackage{tikz}
\usetikzlibrary{shapes}
\usetikzlibrary{quantikz}

\let\originalleft\left
\let\originalright\right
\renewcommand{\left}{\mathopen{}\mathclose\bgroup\originalleft}
\renewcommand{\right}{\aftergroup\egroup\originalright}

\newcommand{\eye}{\operatorname{i}}
\newcommand{\op}[1]{\ensuremath{#1}}%\newcommand{\op}[1]{\ensuremath{\hat{#1}}}

\newcommand{\trace}{\ensuremath{\mathrm{tr}}}
\newcommand{\transpose}{\ensuremath{\mathrm{T}}}
\newcommand{\textgate}[1]{\text{\textsc{#1}}}

\newcommand{\e}{\operatorname{e}}
\newcommand{\abs}[1]{\left| #1 \right|}

\renewcommand{\vec}[1]{\ensuremath{\mathbf{#1}}} % for vectors
\newcommand{\Reals}{\mathds{R}}

\newcommand{\Integers}{\mathds{Z}}

\newcommand{\CR}{\mathrm{CR}}
\newcommand{\CV}{\mathrm{CV}}

\newcommand{\StateCR}{\ensuremath{\op{\rho}}}
\newcommand{\StateCV}{\ensuremath{\op{\tau}}}

\newcommand{\MapDCTCsCR}{\ensuremath{\vec{D}}}

\newcommand{\MapDCTCsCV}{\ensuremath{\vec{d}}}
\newcommand{\MapPCTCsCR}{\ensuremath{\vec{P}}}

\newcommand{\NumberLevels}{\ensuremath{N}}

\newcommand{\Unitary}{\ensuremath{\op{U}}}

\newcommand{\Identity}{\ensuremath{\op{\mathds{1}}}}

\newcommand{\Swap}{\ensuremath{\op{S}}}

\newcommand{\CNOT}[2]{\ensuremath{C\op{X}}^{ #1 , #2 }}
\newcommand{\SUM}{\ensuremath{\op{\Sigma}}}
\newcommand{\CSUM}[2]{\ensuremath{C\op{\Sigma}}^{ #1 , #2 }}
\newcommand{\Pauli}{\ensuremath{\op{\sigma}}}
\newcommand{\PauliGeneral}{\ensuremath{\op{\lambda}}}

\newcommand{\Bell}{\ensuremath{\Phi}}

\newcommand{\ParameterFree}{\ensuremath{g}}

\newcommand{\ParameterPower}{\ensuremath{p}}

\newcommand{\EigenValue}{\ensuremath{\xi}}
\newcommand{\Probe}{\ensuremath{\chi}}
\newcommand{\StateProbe}{\ensuremath{\op{\omega}}}
\newcommand{\StateIntermediate}{\ensuremath{\op{\kappa}}}
\newcommand{\Strength}{\ensuremath{\epsilon}}

\newcommand{\Expectation}{\ensuremath{r}}
\newcommand{\UnitaryTotal}{\ensuremath{\op{W}}}
\newcommand{\UnitaryTomography}{\ensuremath{\op{T}}}
\newcommand{\UnitaryInteraction}{\ensuremath{\op{U}}}
\newcommand{\UnitaryExpansion}{\ensuremath{\op{A}}}
\newcommand{\UnitaryBasis}{\ensuremath{\op{V}}}
\newcommand{\UnitaryHadamard}{\ensuremath{\op{H}}}
\newcommand{\UnitaryPhase}{\ensuremath{\op{R}}}

\newcommand{\SpaceHilbert}{\ensuremath{\mathcal{H}}}

\newcommand{\SpaceLinear}{\ensuremath{\mathcal{L}}}

\newcommand{\via}{via} %through.
\newcommand{\ie}{i.e.} %Literally.
\newcommand{\eg}{e.g.} %For example.
\newcommand{\secn}{Sec.$\!$}

\newcommand{\fig}{Fig.$\!$}
\newcommand{\eqn}{Eq.$\!$}

\newcommand{\rfe}{Ref.$\!$}
\newcommand{\rfs}{Refs.$\!$}
\newcommand{\app}{Appendix}

\begin{document}

\title{Quantum state tomography on closed timelike curves using weak measurements}

\author{Lachlan G. Bishop}
\email{lachlan.bishop@uq.net.au}
\affiliation{%
	School of Mathematics and Physics, The University of Queensland, St. Lucia, Queensland 4072, Australia
}%
\author{Fabio Costa}
\affiliation{%
	Nordita, Stockholm University and KTH Royal Institute of Technology, Hannes Alfv{\'e}ns v{\"a}g 12, SE-106 91 Stockholm, Sweden
}%
\affiliation{%
	School of Mathematics and Physics, The University of Queensland, St. Lucia, Queensland 4072, Australia
}%
\author{Timothy C. Ralph}
\affiliation{%
	School of Mathematics and Physics, The University of Queensland, St. Lucia, Queensland 4072, Australia
}%

\date{\today}

\begin{abstract}
	Any given prescription of quantum time travel necessarily endows a Hilbert space to the chronology-violating (CV) system on the closed timelike curve (CTC). However, under the two foremost models, Deutsch's prescription (D-CTCs) and postselected teleportation (P-CTCs), the CV system is treated very differently: D-CTCs assign a definite form to the state on this system, while P-CTCs do not. To further explore this distinction, we present a methodology by which an operational notion of state may be assigned to their respective CV systems. This is accomplished {\via} a conjunction of state tomography and weak measurements, with the latter being essential in leaving any notions of self-consistency intact. With this technique, we are able to verify the predictions of D-CTCs and, perhaps more significantly, operationally assign a state to the system on the P-CTC. We show that, for any given combination of chronology-respecting input and unitary interaction, it is always possible to recover the unique state on the P-CTC, and we provide a few specific examples in the context of select archetypal temporal paradoxes. We also demonstrate how this state may be derived from analysis of the P-CTC prescription itself, and we explore how it compares to its counterpart in the CV state predicted by D-CTCs.
\end{abstract}

\maketitle
%\tableofcontents

\section{Introduction}\label{sec:introduction}

At first, the concept of time travel and its physical consequences seemed to be fraught with issues regarding the resolutions to seemingly paradoxical scenarios \cite{friedman_cauchy_1990,echeverria_billiard_1991,lossev_jinn_1992,novikov_analysis_1989-1,novikov_time_1992,politzer_simple_1992,mensky_three-dimensional_1996}, sparking a wave of research on causality violation and chronology protection \cite{hawking_chronology_1992,ori_causality_1994,friedman_topological_1997,friedman_topological_2008}. Despite nonrelativistc quantum-mechanical treatments of time travel suffering from various problems \cite{goldwirth_quantum_1994, hartle_unitarity_1994, hawking_quantum_1995, fewster_classical_1996, brun_computers_2003, aaronson_closed_2009, brun_computers_2003, bacon_quantum_2004, aaronson_quantum_2005, brun_localized_2009, brun_perfect_2012, ahn_quantum-state_2013, brun_quantum_2013, ghosh_quantum_2018}, such approaches are generally considered to be more well-equipped for analysing the indeterministic dynamics of physical systems interacting near closed timelike curves (CTCs, known colloquially as `time machines') compared to the traditionally deterministic methods of classical mechanics. Indeed, one of the more interesting facets of such quantum work is the fact that there exists more than one way by which temporal paradoxes may be resolved, the two foremost being Deutsch's prescription (D-CTCs) \cite{deutsch_quantum_1991} and postselected teleportation (P-CTCs) \cite{lloyd_quantum_2011,lloyd_closed_2011}.

A striking observation in a comparison of the two prescriptions is the sheer incompatibility between them. Of their many differences, perhaps the most perplexing of all is that one cannot assign a state to the system on the P-CTC \cite{lloyd_quantum_2011}, while the determinability of the state on the D-CTC is crucial to the success of the theory. For the former, provided all boundary conditions in the path-integral picture \cite{politzer_path_1994} are satisfied by the chronology-violating (CV) state (on the P-CTC), then the ability of the P-CTC prescription to resolve any given time-travel paradox is entirely independent of, and agnostic to, this state. The D-CTC model on the other hand relies fundamentally on the CV state being definable, as it is upon this object which the notion of temporal self-consistency is imposed, with the chronology-respecting (CR) output state being directly dependent on it.

Although the state on the P-CTC is ordinarily unassignable, the P-CTC itself is the conduit by which information is necessarily transported to the past. In principle, this means that some CV state exists physically, and so must admit a definite form. It is therefore natural to ponder exactly what state the quantum system on the P-CTC assumes for any given scenario. Similarly, despite the state on the D-CTC being designated explicitly by its prescription, thus far there exists no technique for actually verifying its form, although an experimental simulation has tested the consistency of the prescription \cite{ringbauer_experimental_2014}. A methodology which provides the ability to investigate the CV states in quantum prescriptions of time travel would therefore be both useful and enlightening. This is the thought which motivates our work here, and the following questions serve to focus our attention:
\begin{enumerate}[label=(\roman*),itemsep=0.015cm]
	\item In general, is it possible to assign an operational definition to the trapped CV state on a CTC?
\end{enumerate}
If so, then:
\begin{enumerate}[label=(\roman*),itemsep=0.015cm]
	\setcounter{enumi}{1}
	\item Can an operational state corresponding to the usual CV state for D-CTCs be constructed?
	\item Is there a similarly determinable operational CV state for P-CTCs?
	\item Does the so-defined trapped P-CTCs CV state obey a self-consistency condition, and how does it otherwise relate to its D-CTCs counterpart?
\end{enumerate}
Admittedly, question (iii) is rather loaded, as it presupposes that there is not a simple yet rigorous way by which a CV state can be quantified through analysis of P-CTCs. As we shall see, there is in fact such a result (at least for qubits), and although its accuracy is not immediately apparent, it can be rigorously proved to be correct. This serves as a segue into the main methodology of this work.

To answer these questions, we employ the well-established technique of \emph{quantum state tomography} \cite{james_measurement_2001, thew_qudit_2002, mauro_dariano_quantum_2003, dariano_2_2004, nielsen_quantum_2010}. This essentially consists of the determination of an unknown quantum state via measurements of it with respect to all elements of an appropriate \emph{tomographically} or \emph{informationally complete basis} (set of observables). Examples of such bases include the Pauli matrices (plus the identity matrix) for qubits, and the Gell-Mann matrices (also with the identity) for qutrits. Once the state has been measured exhaustively in the appropriate basis, the resulting statistics (expectation values) allow for determination of the Bloch vector, which is equivalent to precise identification of the associated (and previously unknown) state.

At first glance, this could conceivably allow for the determination of the state of the time-travelling (CV) system in any given quantum prescription of antichronological time travel. A problem with this methodology however is its incompatibility with the requirement of resolving the state without simultaneously disturbing it ({\ie}, collapsing the wave function). Such a feature is essential, as an ordinary (`strong') measurement would necessarily disturb the CV state, thereby disrupting the relevant self-consistency condition(s) in any given quantum formulation of CTCs (like D-CTCs and P-CTCs). Therefore, in order to measure the state on the CTC without (significantly) perturbing the system, we can use so-called \emph{weak measurements} \cite{aharonov_quantum_2005, aharonov_how_1988, peres_quantum_1989, leggett_comment_1989, aharonov_aharonov_1989, duck_sense_1989, aharonov_properties_1990, knight_weak_1990, story_weak_1991, johansen_weak_2004, vaidman_weak_2009, wu_weak_2009, kedem_modular_2010, wu_weak_2011, nakamura_evaluation_2012, lundeen_procedure_2012, pang_weak_2012, kofman_nonperturbative_2012}, that is, quantum measurements which minimise disturbance by leaving the measured state unaffected (to first order).

% For such a scheme to be successful, the measurements must necessarily be `weak'. The reason for this is simple: an ordinary (`strong') measurement disturbs (collapses) the CV state, thereby disrupting the self-consistency conditions of both D-CTCs and P-CTCs. Therefore, in order to measure the state on the CTC without significantly perturbing the system, weak measurements ({\ie}, those which leave the unknown state unaffected to first order) provide the obvious solution.

The idea of performing quantum state tomography using weak measurements is of course not novel; weak-measurement tomography has been used with great success in a plethora of past experimental applications \cite{knight_weak_1990, hosten_observation_2008, lundeen_experimental_2009, dixon_ultrasensitive_2009, kim_reversing_2009, cho_weak_2010, goggin_violation_2011, zilberberg_charge_2011, lundeen_direct_2011, kocsis_observing_2011, feizpour_amplifying_2011, kim_protecting_2012, rozema_violation_2012, vijay_stabilizing_2012, hatridge_quantum_2013, salvail_full_2013, groen_partial-measurement_2013, malik_direct_2014, blok_manipulating_2014, magana-loaiza_amplification_2014, denkmayr_observation_2014, mirhosseini_compressive_2014, shi_scan-free_2015, mahler_experimental_2016, thekkadath_direct_2016, piacentini_measuring_2016, hallaji_weak-value_2017, kim_direct_2018, nirala_measuring_2019}. Our work here however is the first instance in which time-travelling quantum (qubit) states on CTCs have been quantified theoretically in such a manner.

In {\secn} \ref{sec:background}, we begin our exposition of time-travelling quantum states (\ref{sec:prescriptions}) with brief descriptions of both D-CTCs (\ref{sec:D-CTCs}) and P-CTCs (\ref{sec:P-CTCs}). This is followed with a comprehensive description of our weak-measurement tomography scheme in {\secn} \ref{sec:tomography}. Note that for simplicity, we restrict our procedure to qubits, as the complexity of tomography rapidly increases with the dimensionality of the state's Hilbert space \cite{james_measurement_2001, thew_qudit_2002}. With this in mind, we then demonstrate in {\secn} \ref{sec:models} how both prescriptions (\ref{sec:tomography_D-CTCs} for D-CTCs, \ref{sec:tomography_P-CTCs} for P-CTCs) remain undisturbed under weak measurements.

It is in {\secn} \ref{sec:tomography_P-CTCs} where we present our main findings regarding the formerly unassignable P-CTC CV state. Interestingly, we discover how a state (recovered from our weak-tomography scheme) can be assigned to the system on the P-CTC for \emph{every} given combination of interaction and input state. Furthermore, this state turns out to be manifestly unique (in contrast to its D-CTC counterpart), but this is perhaps unsurprising due to the P-CTC solution being likewise unique. We also show in {\secn} \ref{sec:P-CTCs_CV} how a crude, so-called `na\"{i}ve' computation of the P-CTC CV state from the P-CTC prescription itself may be obtained, and how this state exactly matches the tomographical state, giving a much simpler view to the entire picture.

To put these results into a more specific context, we apply our tomography procedure to a select group of archetypal time-travel problems ({\secn} \ref{sec:results}), including the grandfather (\ref{sec:grandfather}) and unproven theorem (\ref{sec:unproven}) paradoxes. Perhaps surprisingly, for most of the examples, the D-CTC CV state is unique and exactly matches the P-CTC state. However, in the case of the unproven theorem paradox, the D-CTC state is nonunique, possessing parametrised solutions. In contrast, the P-CTC CV state is unique and corresponds to the maximally entropic version of its D-CTC counterpart. We discuss the general relationship between the states further in {\secn} \ref{sec:P-CTC_fixed_points}, wherein we conclude in general that the P-CTC CV state is a D-CTC solution only in the cases of certain interactions that satisfy a derivable constraint. We finish with concluding remarks in {\secn} \ref{sec:conclusion}.

\section{Background}\label{sec:background}

\subsection{Quantum states on CTCs}\label{sec:prescriptions}

\subsubsection{Deutsch's prescription}\label{sec:D-CTCs}

We will begin with an outline of Deutsch's prescription (D-CTCs) \cite{deutsch_quantum_1991}, because it is perhaps the most accessible description of quantum time travel. In this theory, time-travel paradoxes are resolved by imposing self-consistency directly onto the chronology-violating state. An important point is that the local density operator of the system on the D-CTC is taken as a fundamental object, meaning that one is able to (and necessarily must) assign a definite state to this system. An operational meaning for this trapped state is missing however, and part of our work in {\secn} \ref{sec:tomography_D-CTCs} is to provide one.

\begin{figure}[h]
	\includegraphics[scale=1]{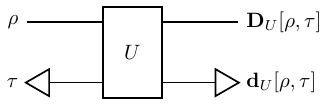}
	\caption[D-CTC circuit]{\label{fig:circuit_D-CTCs}A circuit diagram of Deutsch's prescription (D-CTCs). With time progressing from left to right, the upper and lower modes correspond to the CR and CV Hilbert spaces respectively. The interaction between these is described by the unitary gate $\UnitaryInteraction$, and the triangle caps at each end of the CV mode represent the past (left) and future (right) openings of the time machine.}
\end{figure}

Figure \ref{fig:circuit_D-CTCs} depicts a diagrammatic quantum circuit formulation of the prescription. Let $\StateCR \in \SpaceLinear(\SpaceHilbert_\mathrm{CR})$ be the chronology-respecting (CR) input state, and $\StateCV \in \SpaceLinear(\SpaceHilbert_\mathrm{CV})$ be the chronology-violating (CV) state on the D-CTC. Here, $\SpaceLinear(\SpaceHilbert)$ is the space of linear operators on the Hilbert space $\SpaceHilbert$. Beginning with the input bipartite product state $\StateCR \otimes \StateCV$, the interaction between the CR and CV states {\via} the unitary $\UnitaryInteraction$ is encoded by the map
\begin{equation}
	\StateCR \otimes \StateCV \rightarrow \Unitary\left(\StateCR\otimes\StateCV\right)\Unitary^\dagger. \label{eq:D-CTCs_map}
\end{equation}
By defining the CV map,
\begin{equation}
	\MapDCTCsCV_{\UnitaryInteraction}\left[\StateCR,\StateCV\right] \equiv \trace_\CR\left[\Unitary\left(\StateCR\otimes\StateCV\right)\Unitary^\dagger\right], \label{eq:D-CTCs_CV}
\end{equation}
where the partial trace is over the CR subsystem, temporal self-consistency of the state on the CTC is imposed by demanding that the past state $\StateCV$ be equal to its future (post-interaction) counterpart, that is,
\begin{equation}
	\StateCV = \MapDCTCsCV_{\UnitaryInteraction}\left[\StateCR,\StateCV\right]. \label{eq:D-CTCs_CV_fixed}
\end{equation}
A solution (fixed point) of this map is interpreted as being the state which is trapped on the D-CTC, and has the corresponding CR output solution state given by
\begin{equation}
	\MapDCTCsCR_{\UnitaryInteraction}\left[\StateCR,\StateCV\right] \equiv \trace_\CV\left[\Unitary\left(\StateCR\otimes\StateCV\right)\Unitary^\dagger\right]. \label{eq:D-CTCs_CR}
\end{equation}
If the context is clear, we will express (\ref{eq:D-CTCs_CR}) as $\StateCR_\mathrm{D}\equiv\MapDCTCsCR_{\UnitaryInteraction}\left[\StateCR,\StateCV\right]$ for brevity.

One of the most significant characteristics of Deutch's prescription is that, depending on the interaction $\UnitaryInteraction$, the self-consistency condition (\ref{eq:D-CTCs_CV_fixed}) may be satisfied by more than one solution. This characteristic, known as the \emph{uniqueness ambiguity} in the context of D-CTCs, may be interpreted as an indication that the theory is incomplete. Deutsch's argument against this is the assertion that in the case of a multiplicity of the fixed points, the `correct' state is the one which possesses the most entropy \cite{deutsch_quantum_1991}. To some, this proposed resolution is rather unsatisfactory, and so the D-CTC equivalent circuit picture (ECP) \cite{ralph_information_2010,ralph_reply_2011,pienaar_quantum_2011,ralph_relativistic_2012,dong_ralphs_2017} was formulated to provide a more rigorous solution.

In the ECP, the form of the D-CTC CV state becomes intuitive when it is interpreted as being the end-state result of evolving some initial `seed' state $\StateCV^{(0)}$ through the circuit iteratively until a unique fixed point is reached. In general, the nature of the $\StateCV^{(0)}$ can be seen to parametrise the CTC ({\eg}, in terms of the energy trapped when it was formed). At each step, an identical copy of the input $\StateCR$ is prepared on the CR mode, which is then traced out following interaction with the CV mode. By recursion, we may write this mathematically as
\begin{equation}
	\StateCV^{(n+1)} = \MapDCTCsCV_{\UnitaryInteraction}\bigl[\StateCR,\StateCV^{(n)}\bigr] \equiv \trace_\mathrm{CR}\bigl[\UnitaryInteraction\bigl(\StateCR\otimes\StateCV^{(n)}\bigr)\UnitaryInteraction^\dagger\bigr] \label{eq:ECP_recursion}
\end{equation}
which, in the limit as $n \rightarrow \infty$, converges to the fixed point (denoted by $\StateCV^{(\infty)}$) uniquely corresponding to the given seed state. For many specific interactions $\UnitaryInteraction$, this is the same state as predicted by Deutsch, but there are other scenarios in which the distinction between Deutsch's rule of maximum entropy and the ECP is significant (see {\rfe} \cite{dong_ralphs_2017} for discussion). Regardless, the ECP prescription of the seed state $\StateCV^{(0)}$ together with the recursive map (\ref{eq:ECP_recursion}) forms a more satisfactory resolution to the uniqueness ambiguity than Deutsch's entropy rule because it is the consequence of a simple postulate, not an educated conjecture.

\subsubsection{Postselected teleportation}\label{sec:P-CTCs}

\begin{figure}[h]
	\includegraphics[scale=1]{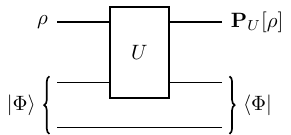}
	\caption[P-CTC circuit]{\label{fig:circuit_P-CTCs}A circuit diagram of the postselected teleportation prescription (P-CTCs). Here, the CV subsystem consists of the lower two modes, upon which the maximally entangled state is preselected (in the past as $\ket{\Bell}$) and postselected (in the future as $\bra{\Bell}$).}
\end{figure}

In contrast to D-CTCs, the postselected teleportation prescription (P-CTCs) \cite{lloyd_quantum_2011, lloyd_closed_2011} provides self-consistent resolutions to quantum time-travel paradoxes by way of quantum teleportation. This is achieved by first replacing the CV state $\StateCV \in \SpaceLinear(\SpaceHilbert_\mathrm{CV})$ with a maximally entangled state $\ket{\Bell} \in \SpaceHilbert_\mathrm{CV} \otimes \SpaceHilbert_\mathrm{T}$, where $\SpaceHilbert_\mathrm{T}$ is the \emph{teleportation} subsystem. See {\fig} \ref{fig:circuit_P-CTCs} for a pictorial representation of the P-CTC protocol. With this, the evolution of the product input state $\StateCR \otimes \ket{\Bell}\bra{\Bell}$ under the unitary $\UnitaryInteraction$ is then given by
\begin{equation}
	\StateCR \otimes \ket{\Bell}\bra{\Bell} \rightarrow \bigl(\UnitaryInteraction \otimes \Identity\bigr)\bigl(\StateCR \otimes \ket{\Bell}\bra{\Bell}\bigr)\bigl(\UnitaryInteraction^\dagger \otimes \Identity\bigr). \label{eq:P-CTCs_evolved}
\end{equation}
In accordance with the prescription, subsequent postselection against the same entangled state results in a quantum channel to the past (on the subsystem $\SpaceHilbert_\mathrm{T}$). Mathematically, the evolved CR state becomes
\begin{align}
	\StateCR_\mathrm{P} &\propto \bigl(\Identity \otimes \bra{\Bell}\bigr)\bigl(\UnitaryInteraction \otimes \Identity\bigr)\bigl(\StateCR \otimes \ket{\Bell}\bra{\Bell}\bigr)\bigl(\UnitaryInteraction^\dagger \otimes \Identity\bigr)\bigl(\Identity \otimes \ket{\Bell}\bigr) \nonumber\\
	&= \trace_\mathrm{CV}[\UnitaryInteraction]\,\StateCR\,\trace_\mathrm{CV}[\UnitaryInteraction^\dagger], \label{eq:P-CTCs_CR_unnormalised}
\end{align}
where we computed
\begin{equation}
	\bigl(\Identity \otimes \bra{\Bell}\bigr)\bigl(\UnitaryInteraction \otimes \Identity\bigr)\bigl(\Identity \otimes \ket{\Bell}\bigr) \propto \trace_\mathrm{CV}[\UnitaryInteraction].
\end{equation}
The P-CTCs evolution of the input state $\StateCR$ is then given by the normalised form of the nonunitary map (\ref{eq:P-CTCs_CR_unnormalised}), that is,
\begin{equation}
	\MapPCTCsCR_{\UnitaryInteraction}\left[\StateCR\right] = \frac{\trace_\mathrm{CV}[\UnitaryInteraction]\,\StateCR\,\trace_\mathrm{CV}[\UnitaryInteraction^\dagger]}{\trace\bigl\{\trace_\mathrm{CV}[\UnitaryInteraction]\,\StateCR\,\trace_\mathrm{CV}[\UnitaryInteraction^\dagger]\bigr\}}. \label{eq:P-CTCs_CR}
\end{equation}
For simplicity, we will often denote this using the notation $\StateCR_\mathrm{P} \equiv \MapPCTCsCR_{\UnitaryInteraction}\left[\StateCR\right]$ when the context ({\ie}, the given $\UnitaryInteraction$ and $\StateCR$) is clear.

Perhaps the most significant characteristic of P-CTCs, at least in relevance to our interests in this work, is its agnosticism regarding the state on the P-CTC. In other words, to resolve time-travel paradoxes, one need not define a CV state, nor does the output state given by (\ref{eq:P-CTCs_CR}) depend in any way on such a state. We argue however that this is rather unsatisfying, given that the particle(s) traversing the P-CTC must necessarily be describable by a state in transit on some superposition of paths in spacetime.%In fact, the very discussion of a P-CTC `trapped' state may be misleading, since the teleportation generated by the Bell state implies that the physical state of the system can only be completely characterised by an entangled bipartite system.

\subsection{State tomography {\via} weak measurements}\label{sec:tomography}

We present here our methodology for qubit weak-measurement tomography in the specific context of our control-target setup. Note that while most treatments consider weak measurements characterised by weakly coupled interactions (that is, unitary operators close to identity), our formalism is based on {\rfs} \cite{pryde_measuring_2004,pryde_measurement_2005}, in which a near eigenstate of the \textgate{not} (Pauli-X) gate is perturbed `weakly' by a control state {\via} the standard controlled-\textgate{not} interaction. See, {\eg}, \cite{wu_weak_2009, kofman_nonperturbative_2012, wu_state_2013, svensson_pedagogical_2013, tamir_introduction_2013, dressel_colloquium_2014} for pedagogical exposition of weak measurements and the associated weak values, and {\rfs} \cite{story_weak_1991, wu_state_2013, kim_direct_2018, botero_weak_2018} for good examples of weak-measurement tomography.

%%% We will show how such a form can be used to reconstruct an unknown state $\StateCR$ when the two are coupled together with a controlled-\textgate{NOT} gate

At the heart of our analysis is an ancillary qubit (the `target' or `probe'), which in the $z$-basis $\left\{\ket{0},\ket{1}\right\}$ may be expressed as
\begin{equation}
	\ket{\Probe} = \sqrt{\frac{1+\Strength}{2}}\ket{0} + \sqrt{\frac{1-\Strength}{2}}\ket{1}, \quad 0 \leq \Strength \leq 1. \label{eq:probe}
\end{equation}
Its function is to transform in response to the state of an unknown system $\StateCV$, which is achieved by coupling the two states together. A direct measurement of the ancilla subsequently yields information regarding $\StateCV$, and if the interaction is sufficiently weak, then this state remains undisturbed. Note that of course, a quantum state cannot be fully characterised by performing just a single measurement. The best we can accomplish for qubits, at least using weak measurements, is to infer the expectation value of the unknown system along one of the three axes of the Bloch sphere. The original state $\StateCV$ may then be completely reconstructed from the combined statistics, by way of the determination of the state's Bloch vector.

Perhaps the simplest coupling between an unknown state and the probe that will permit our tomography scheme is the controlled-\textgate{not} gate, which we write as
\begin{equation}
	\CNOT{1}{2} \equiv {\ket{0}\bra{0}}^1 \otimes \Identity^2 + {\ket{1}\bra{1}}^1 \otimes {\Pauli_x}^2.
\end{equation}
{Note that the superscripts indicate the subsystems on which the operators act: here, the control is on the first while the \textgate{not} acts on the second.} As we wish to measure along each of three Bloch axes, this can be combined with basis transformation gates $\UnitaryBasis^{(\dagger)}_k$, which yields a system-probe unitary interaction
\begin{align}
	\UnitaryTomography &= \bigl(\UnitaryBasis^{\dagger}_k \otimes \Identity\bigr) \cdot \CNOT{1}{2} \cdot \bigl(\UnitaryBasis_k \otimes \Identity\bigr) \nonumber\\
	&= \ket{0_k}\bra{0_k} \otimes \Identity + \ket{1_k}\bra{1_k} \otimes \Pauli_x. \label{eq:unitary_measurement}
\end{align}

\begin{figure}[b]
	\includegraphics[scale=1]{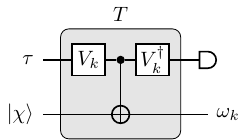}
	\caption[Weak-measurement tomography circuit]{\label{fig:circuit_tomography}A circuit illustration of our weak-measurement scheme. The lower mode is the ancillary subsystem, upon which the probe state $\ket{\Probe}$ interacts with the unknown state $\StateCV$ subsystem {\via} a controlled-\textgate{not}, thereby evolving to the form $\StateProbe_k$, while the unknown system becomes $\StateCV'$.} % The meter at the future end of the ancilla's mode represents a strong, post-interaction measurement.}
\end{figure}

This is depicted diagrammatically as a quantum circuit in {\fig} \ref{fig:circuit_tomography}, where we denote the evolved ancilla by $\StateProbe_k$. Here, the aforementioned basis gates $\UnitaryBasis^{(\dagger)}_k$ (where $k=1,2,3$) represent transformations which map the $z$-basis eigenstates $\left\{\ket{0},\ket{1}\right\}$ into either the $x$- ($k=1$), $y$- ($k=2$) or $z$- ($k=3$) bases. Specifically, we write
\begin{subequations}
\begin{align}
	\UnitaryBasis^\dagger_k  \ket{0} &= \ket{0_k}, \\
	\UnitaryBasis^\dagger_k  \ket{1} &= \ket{1_k},
\end{align}\label{eq:eigenstates}
\end{subequations}
where $\left\{\ket{0_k},\ket{1_k}\right\}$ are eigenstates of the three bases of our qubit Hilbert space. This means that, given the Pauli matrices,
\begin{subequations}
\begin{align}
	\op{\Pauli}_1 = \op{\Pauli}_x &\equiv \ket{0}\bra{1} + \ket{1}\bra{0}, \\
	\op{\Pauli}_2 = \op{\Pauli}_y &\equiv -\eye\bigl(\ket{0}\bra{1} - \ket{1}\bra{0}\bigr), \\
	\op{\Pauli}_3 = \op{\Pauli}_z &\equiv \ket{0}\bra{0} - \ket{1}\bra{1},
\end{align}
\end{subequations}
the states defined by (\ref{eq:eigenstates}) necessarily satisfy
\begin{subequations}
\begin{align}
	\Pauli_k  \ket{0_k} &= \ket{0_k}, \\
	\Pauli_k  \ket{1_k} &= -\ket{1_k}.
\end{align}
\end{subequations}
For this to be fulfilled, we must have
\begin{subequations}
\begin{align}
	\UnitaryBasis_1 &= \UnitaryHadamard, \\
	\UnitaryBasis_2 &= \UnitaryHadamard \UnitaryPhase^\dagger, \\
	\UnitaryBasis_3 &= \Identity,
\end{align}\label{eq:transformations}
\end{subequations}
where
\begin{equation}
	\UnitaryHadamard \equiv \frac{1}{\sqrt{2}} \bigl(\ket{0}\bra{0} + \ket{0}\bra{1} + \ket{1}\bra{0} - \ket{1}\bra{1}\bigr)
\end{equation}
is the Hadamard gate, and
\begin{equation}
	\UnitaryPhase \equiv \ket{0}\bra{0} + \eye\ket{1}\bra{1}
\end{equation}
is the phase shift gate.

With these definitions in mind, a weak measurement can be easily computed in a number of steps. First, given the unknown system $\StateCV$, the ancilla $\ket{\Probe}$, and our weak-measurement unitary (\ref{eq:unitary_measurement}), the evolution of the product state $\StateCV\otimes\ket{\Probe}\bra{\Probe}$ is the standard unitary map
\begin{align}
	\StateCV\otimes\ket{\Probe}\bra{\Probe} \rightarrow \; & \UnitaryTomography\bigl(\StateCV\otimes\ket{\Probe}\bra{\Probe}\bigr)\UnitaryTomography^\dagger \nonumber\\
	%= \; & \aket{0_k}\abra{0_k} \StateCV \aket{0_k}\abra{0_k} \otimes \aket{\Probe}\abra{\Probe} \nonumber\\
	%&+ \aket{0_k}\abra{0_k} \StateCV \aket{1_k}\abra{1_k} \otimes \aket{\Probe}\abra{\Probe}\Pauli_x \nonumber\\
	%&+ \aket{1_k}\abra{1_k} \StateCV \aket{0_k}\abra{0_k} \otimes \Pauli_x\aket{\Probe}\abra{\Probe} \nonumber\\
	%&+ \aket{1_k}\abra{1_k} \StateCV \aket{1_k}\abra{1_k} \otimes \Pauli_x\aket{\Probe}\abra{\Probe}\Pauli_x \nonumber\\
	= \; & \sum_{n,m=0}^{1} \ket{n_k}\bra{n_k}  \StateCV  \ket{m_k}\bra{m_k} \otimes \Pauli^n_x  \ket{\Probe}\bra{\Probe}  \Pauli_x^{m}. \label{eq:measurement_evolution}
\end{align}
The state of the evolved system $\StateCV'$ after the interaction is computed by performing a partial trace on the probe's Hilbert space,
\begin{align}
	\StateCV' = \; & \trace_{2}\bigl[\UnitaryTomography\bigl(\StateCV\otimes\ket{\Probe}\bra{\Probe}\bigr)\UnitaryTomography^\dagger\bigr] \nonumber\\
	%= \; & \aket{0_k}\abra{0_k} \StateCV \aket{0_k}\abra{0_k} \nonumber\\
	%&+ \abra{\Probe}{\Pauli_x}\aket{\Probe} \aket{0_k}\abra{0_k} \StateCV \aket{1_k}\abra{1_k} \nonumber\\
	%&+ \abra{\Probe}{\Pauli_x}\aket{\Probe} \aket{1_k}\abra{1_k} \StateCV \aket{0_k}\abra{0_k} \nonumber\\
	%&+ \aket{1_k}\abra{1_k} \StateCV \aket{1_k}\abra{1_k}.
	= \; & \sum_{n,m=0}^{1} \bra{\Probe}  \Pauli_x^{m} \Pauli_x^n  \ket{\Probe} \cdot \ket{n_k}\bra{n_k}  \StateCV  \ket{m_k}\bra{m_k}. \label{eq:system_evolved}
\end{align}
To show that the measurement is necessarily weak, thereby leaving this evolved state unperturbed for sufficiently small $\Strength$, we first use the definition of our normalised probe state (\ref{eq:probe}) to compute
\begin{equation}
	\bra{\Probe} \Pauli_x  \ket{\Probe} = \sqrt{1-\Strength^2}. \label{eq:probe_overlap}
\end{equation}
A Taylor series expansion of this lets us write
\begin{equation}
	\sqrt{1-\Strength^2} = 1 - \frac{\Strength^2}{2} + \mathcal{O}(\Strength^4), \label{eq:expansion}
\end{equation}
with which we may express the evolved system (\ref{eq:system_evolved}) as
\begin{align}
	\StateCV' %=\; & \ket{0_k}\bra{0_k} \StateCV \ket{0_k}\bra{0_k} \nonumber\\
	%&+ \aket{0_k}\abra{0_k} \StateCV \aket{1_k}\abra{1_k} \nonumber\\
	%&+ \aket{1_k}\abra{1_k} \StateCV \aket{0_k}\abra{0_k} \nonumber\\
	%&+ \aket{1_k}\abra{1_k} \StateCV \aket{1_k}\abra{1_k} \nonumber\\
	%&- \left(\frac{\Strength^2}{2} - \mathcal{O}(\Strength^4)\right) \left[\aket{0_k}\abra{0_k} \StateCV \aket{1_k}\abra{1_k} + \aket{1_k}\abra{1_k} \StateCV \aket{0_k}\abra{0_k}\right] \nonumber\\
	%=\; & \sum_{n,m=0}^{1} \aket{n_k}\abra{n_k} \StateCV \aket{m_k}\abra{m_k} \nonumber\\
	%&- \left(\frac{\Strength^2}{2} - \mathcal{O}(\Strength^4)\right) \left[\aket{0_k}\abra{0_k} \StateCV \aket{1_k}\abra{1_k} + \aket{1_k}\abra{1_k} \StateCV \aket{0_k}\abra{0_k}\right] \nonumber\\
	=\; & \sum_{n,m=0}^{1} \ket{n_k}\bra{n_k}  \StateCV  \ket{m_k}\bra{m_k} + \mathcal{O}(\Strength^2). \label{eq:system_expansion}
\end{align}
By comparing this to our probe density, which can be expanded as
\begin{align}
	\ket{\Probe}\bra{\Probe} %&= \Identity + \Strength\Pauli_z + \sqrt{1-\Strength^2}\Pauli_x, \nonumber\\
	%&= \ket{+}\bra{+} + \Strength\Pauli_z - \left(\frac{\Strength^2}{2} -\mathcal{O}(\Strength^4)\right)\Pauli_x, \nonumber\\
	&= \ket{+}\bra{+} + \frac{\Strength}{2}\Pauli_z + \mathcal{O}(\Strength^2), \label{eq:probe_density}
\end{align}
it is easy to see that for sufficiently small $\abs{\Strength} \ll 1$ such that $\mathcal{O}(\Strength^2) \approx 0$, the system state (\ref{eq:system_expansion}) approximates its unperturbed form,
\begin{align}
	\StateCV' \approx\; & \sum_{n,m=0}^{1} \ket{n_k}\bra{n_k}  \StateCV  \ket{m_k}\bra{m_k} \nonumber\\
	%=\; & \aket{0_k}\abra{0_k} \StateCV \aket{0_k}\abra{0_k} \nonumber\\
	%&+ \aket{0_k}\abra{0_k} \StateCV \aket{1_k}\abra{1_k} \nonumber\\
	%&+ \aket{1_k}\abra{1_k} \StateCV \aket{0_k}\abra{0_k} \nonumber\\
	%&+ \aket{1_k}\abra{1_k} \StateCV \aket{1_k}\abra{1_k} \nonumber\\
	=\; & \StateCV,
\end{align}
while the probe (\ref{eq:probe_density}), containing a first-order power of $\Strength$, becomes
\begin{equation}
	\ket{\Probe}\bra{\Probe} \approx \ket{+}\bra{+} + \frac{\Strength}{2}\Pauli_z.
\end{equation}
This is exactly what we desire, since if the probe were to become exactly an eigenstate of the Pauli-X operator $\Pauli_x$, {\ie},
\begin{equation}
	\ket{\pm} \equiv \frac{1}{\sqrt{2}}\bigl(\ket{0} \pm \ket{1}\bigr),
\end{equation}
then it would not be perturbed by the controlled-\textgate{not} gate in its coupling (\ref{eq:unitary_measurement}) with the unknown system state. In other words, it would be unable to provide any information regarding $\StateCV$, and would thus be tomographically useless. As however the probe state (\ref{eq:probe_density}) does not assume the form of an eigenstate of $\Pauli_x$ while the system remains unperturbed by the interaction (provided $\Strength$ is sufficiently small), then we can in principle infer the state of $\StateCV$ {\via} strong measurements of the evolved ancilla state $\StateProbe_k$ (for all $k$). This is the meaning of a weak measurement, and we call $\Strength$ the \emph{strength} of the measurement.

Let us now demonstrate how the qubit system $\StateCV$ may be reconstructed by measuring the probe in the $z$ basis. Starting with (\ref{eq:measurement_evolution}), we perform a partial trace on the first subsystem to obtain the evolved probe state,
\begin{align}
	\StateProbe_k &\equiv \trace_{1}\bigl[\UnitaryTomography\bigl(\StateCV\otimes\ket{\Probe}\bra{\Probe}\bigr)\UnitaryTomography^\dagger\bigr] \nonumber\\
	%&= \abra{0_k} \StateCV \aket{0_k} \aket{\Probe}\abra{\Probe} + \abra{1_k} \StateCV \aket{1_k} \Pauli_x\aket{\Probe}\abra{\Probe}\Pauli_x \nonumber\\
	&= \sum_{n=0}^{1} \bra{n_k}  \StateCV  \ket{n_k} \cdot \Pauli^n_x  \ket{\Probe}\bra{\Probe}  \Pauli_x^{n}. \label{eq:state_probe}
\end{align}
Computing the expectation value of $\Pauli_z$ for this form then yields
\begin{align}
	\trace\left[\Pauli_z\StateProbe_k\right] &= \bra{0}  \StateProbe_k  \ket{0} - \bra{1}  \StateProbe_k  \ket{1} \nonumber\\
	&= \Strength \bigl(\bra{0_k}  \StateCV  \ket{0_k} - \bra{1_k}  \StateCV  \ket{1_k}\bigr) \nonumber\\
	&= \Strength \, \trace\left[\Pauli_k \StateCV\right], \label{eq:connection}
\end{align}
where we used the fact that
\begin{equation}
	\Pauli_k = \ket{0_k}\bra{0_k} - \ket{1_k}\bra{1_k}.
\end{equation}
Rescaling by a factor $\tfrac{1}{\Strength}$ then allows us to make the connection between performing a $z$-basis measurement on the probe state (\ref{eq:state_probe}) and inferring the expectation value of $\Pauli_k$ for the state of the unknown system. This is because, up to the factor $\Strength$, the two operations are equivalent. Note that, for brevity, we will write
\begin{equation}
	\Expectation_k \equiv \frac{1}{\Strength}\trace\left[\Pauli_z\StateProbe_k\right] = \trace\left[\Pauli_k \StateCV\right], \label{eq:expectation}
\end{equation}
which is often denoted in the literature as $\big<\Pauli_k\big>$.

If we define $\Pauli_0 \equiv \Identity$, then the set of operators $\{\Pauli_k\}_{k=0}^3$ form a complete basis for the space of complex $2 \times 2$ matrices. Since this coincides with the space of linear operators $\SpaceLinear(\SpaceHilbert)$ on our qubit ($2$-dimensional) Hilbert space $\SpaceHilbert$, then any state $\StateCV \in \SpaceLinear(\SpaceHilbert)$ can accordingly be reconstructed {\via} the linear combination,
\begin{equation}
	\StateCV = \frac{1}{2}\sum_{k=0}^{3} \trace\left[\Pauli_k \StateCV\right] \Pauli_k. \label{eq:reconstruction}
\end{equation}
Since the coefficient for $k = 0$ is fixed by normalisation ({\ie} $\trace\left[\Pauli_0 \tau \right] = 1$), the state is fully determined by the \emph{Bloch vector}, which is defined, for an $\NumberLevels$-dimensional state, as the $(\NumberLevels^2-1)$-dimensional real vector of parameters $\trace\left[\Pauli_k \tau \right]$ (for $k \geq 1$). Accordingly, the linear combination (\ref{eq:reconstruction}) is said to be the \emph{Bloch sphere representation} of the state $\tau$. Thus, since we have shown that we can infer expectation values (\ref{eq:expectation}) for an unknown state by coupling it with a probe {\via} a controlled-\textgate{NOT} and subsequently performing a measurement of the probe in the $z$-basis, then we can characterise the state completely without directly measuring it.

Given this tomography scheme, it is in principle possible to rigorously assign an operational definition to the trapped state on a CTC. This is achieved by simply setting the probe state to interact with the CV mode, thereby allowing reconstruction of the unknown CTC state. Thus, we conclude that our focus question (i) can be answered in the affirmative.

\section{Weak-measurement tomography on CTCs}\label{sec:models}

Of course, the form of the CV state reconstructed from weak measurements could depend completely on the prescription one chooses to work with. In this section, we show how our tomography scheme may be incorporated into each of D-CTCs and P-CTCs, thereby answering both our focus questions (ii) and (iii).

\subsection{Weak-measurement tomography of the state on the D-CTC}\label{sec:tomography_D-CTCs}

\begin{figure}[h]
	\includegraphics[scale=1]{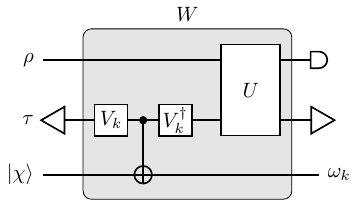}
	\caption[Weak-measurement tomography circuit for D-CTCs]{\label{fig:circuit_tomography_D-CTCs}A circuit of our tomography scheme ({\fig} \ref{fig:circuit_tomography}) incorporated in the D-CTC circuit ({\fig} \ref{fig:circuit_D-CTCs}).}
\end{figure}

To verify our operational methodology, we will perform our weak-measurement tomography scheme on a general D-CTC circuit in order to demonstrate its success in being able to fully characterise the trapped CV state {\via} measurement of the probe alone. Consider, in {\fig} \ref{fig:circuit_tomography_D-CTCs}, the general D-CTC circuit that we will use for this analysis. Note that (without loss of generality) we choose to place the CV-ancilla tomography portion before the CR-CV interaction $\UnitaryInteraction$. This is because, due to the D-CTC fixed-point condition (\ref{eq:D-CTCs_CV}), the CV system assumes the same state both before and after the interaction.

The total unitary corresponding to the circuit depicted in {\fig} \ref{fig:circuit_tomography_D-CTCs} is
\begin{equation}
	\UnitaryTotal = \bigl(\UnitaryInteraction \otimes \Identity\bigr) \cdot \bigl(\Identity \otimes \UnitaryBasis^{\dagger}_k \otimes \Identity\bigr) \cdot \CNOT{2}{3} \cdot \bigl(\Identity \otimes \UnitaryBasis_k \otimes \Identity\bigr),
\end{equation}
with which the evolution of the tripartite input state $\StateCR \otimes \StateCV \otimes \ket{\Probe}\bra{\Probe}$ can be computed to be
\begin{align}
	\StateCR\otimes\StateCV\otimes\ket{\Probe}\bra{\Probe} \rightarrow \; & \UnitaryTotal\bigl(\StateCR\otimes\StateCV\otimes\ket{\Probe}\bra{\Probe}\bigr)\UnitaryTotal^\dagger \nonumber\\
	%= \; & \UnitaryInteraction\left(\StateCR \otimes \aket{0_k}\abra{0_k} \StateCV \aket{0_k}\abra{0_k} \right)\UnitaryInteraction^\dagger \otimes \aket{\Probe}\abra{\Probe} \nonumber\\
	%&+ \UnitaryInteraction\left(\StateCR \otimes \aket{0_k}\abra{0_k} \StateCV \aket{1_k}\abra{1_k} \right)\UnitaryInteraction^\dagger \otimes \aket{\Probe}\abra{\Probe}\Pauli_x \nonumber\\
	%&+ \UnitaryInteraction\left(\StateCR \otimes \aket{1_k}\abra{1_k} \StateCV \aket{0_k}\abra{0_k} \right)\UnitaryInteraction^\dagger \otimes \Pauli_x\aket{\Probe}\abra{\Probe} \nonumber\\
	%&+ \UnitaryInteraction\left(\StateCR \otimes \aket{1_k}\abra{1_k} \StateCV \aket{1_k}\abra{1_k} \right)\UnitaryInteraction^\dagger \otimes \Pauli_x\aket{\Probe}\abra{\Probe}\Pauli_x \nonumber\\
	= \; & \sum_{n,m=0}^{1} \UnitaryInteraction\bigl(\StateCR \otimes \ket{n_k}\bra{n_k}  \StateCV  \ket{m_k}\bra{m_k} \bigr)\UnitaryInteraction^\dagger \nonumber\\
	& \qquad \quad \otimes \Pauli^n_x  \ket{\Probe}\bra{\Probe}  \Pauli_x^{m}. \label{eq:D-CTCs_general}
\end{align}
If we trace out the probe, we are left with the evolution of the CR and CV modes,
\begin{align}
	\StateCR \otimes \StateCV \rightarrow\; & \trace_3\bigl[\UnitaryTotal\bigl(\StateCR\otimes\StateCV\otimes\ket{\Probe}\bra{\Probe}\bigr)\UnitaryTotal^\dagger\bigr] \nonumber\\
	=\; & \sum_{n,m=0}^{1} \bra{\Probe}  \Pauli_x^{m} \Pauli^n_x  \ket{\Probe} \UnitaryInteraction\bigl(\StateCR \otimes \ket{n_k}\bra{n_k}  \StateCV  \ket{m_k}\bra{m_k} \bigl)\UnitaryInteraction^\dagger \nonumber\\
	%=\; & \UnitaryInteraction\left(\StateCR \otimes \aket{0_k}\abra{0_k} \StateCV \aket{0_k}\abra{0_k} \right)\UnitaryInteraction^\dagger \nonumber\\
	%&+ \abra{\Probe}\Pauli_x\aket{\Probe} \UnitaryInteraction\left(\StateCR \otimes \aket{0_k}\abra{0_k} \StateCV \aket{1_k}\abra{1_k} \right)\UnitaryInteraction^\dagger \nonumber\\
	%&+ \abra{\Probe}\Pauli_x\aket{\Probe} \UnitaryInteraction\left(\StateCR \otimes \aket{1_k}\abra{1_k} \StateCV \aket{0_k}\abra{0_k} \right)\UnitaryInteraction^\dagger  \nonumber\\
	%&+ \UnitaryInteraction\left(\StateCR \otimes \aket{1_k}\abra{1_k} \StateCV \aket{1_k}\abra{1_k} \right)\UnitaryInteraction^\dagger \nonumber\\
	%=\; & \UnitaryInteraction\left(\StateCR \otimes \StateCV \right)\UnitaryInteraction^\dagger \nonumber\\
	%&- \left(\frac{\Strength^2}{2} - \mathcal{O}(\Strength^4)\right)\bigl[ \UnitaryInteraction\left(\StateCR \otimes \aket{0_k}\abra{0_k} \StateCV \aket{1_k}\abra{1_k} \right)\UnitaryInteraction^\dagger \nonumber\\
	%&\qquad \qquad \qquad \qquad + \UnitaryInteraction\left(\StateCR \otimes \aket{1_k}\abra{1_k} \StateCV \aket{0_k}\abra{0_k} \right)\UnitaryInteraction^\dagger \bigr] \nonumber\\
	=\; & \UnitaryInteraction(\StateCR \otimes \StateCV)\UnitaryInteraction^\dagger + \mathcal{O}(\Strength^2)
\end{align}
where we used the expansion (\ref{eq:probe_density}). Evidently, for small $\Strength$, everything but the first term vanishes, leaving us with the ordinary D-CTC evolution as described by (\ref{eq:D-CTCs_map}). This can be verified by first computing the CV map (\ref{eq:D-CTCs_CV}) associated with this evolution,
\begin{equation}
	\MapDCTCsCV_{\UnitaryInteraction}\left[\StateCR,\StateCV\right] \equiv \trace_1 \left[\Unitary\left(\StateCR\otimes\StateCV\right)\Unitary^\dagger\right] + \mathcal{O}(\Strength^2),
\end{equation}
with which it is easy to see that, from (\ref{eq:D-CTCs_CV_fixed}), the D-CTC fixed points ({\ie}, CV solutions) are obtained {\via}
\begin{equation}
	\StateCV = \trace_1 \left[\Unitary\left(\StateCR\otimes\StateCV\right)\Unitary^\dagger\right] + \mathcal{O}(\Strength^2).
\end{equation}
This demonstrates that the weak measurement does not influence the CV state, which in turn means that the associated CR state, computable using (\ref{eq:D-CTCs_CR}),
\begin{equation}
	\MapDCTCsCR_{\UnitaryInteraction}\left[\StateCR,\StateCV\right] \equiv \trace_2 \left[\Unitary\left(\StateCR\otimes\StateCV\right)\Unitary^\dagger\right] + \mathcal{O}(\Strength^2).
\end{equation}
is likewise unaffected for sufficiently small $\Strength$.

In (\ref{eq:D-CTCs_general}), performing a partial trace on the CR and CV modes instead yields the probe evolution,
\begin{align}
	\StateProbe_k =\; & \trace_{1,2}\bigl[\UnitaryTotal\bigl(\StateCR\otimes\StateCV\otimes\ket{\Probe}\bra{\Probe}\bigr)\UnitaryTotal^\dagger\bigr] \nonumber\\
	%=\; & \trace\bigl[\UnitaryInteraction\left(\StateCR \otimes \aket{0_k}\abra{0_k} \StateCV \aket{0_k}\abra{0_k} \right)\UnitaryInteraction^\dagger\bigr] \aket{\Probe}\abra{\Probe} \nonumber\\
	%&+ \trace\bigl[\UnitaryInteraction\left(\StateCR \otimes \aket{0_k}\abra{0_k} \StateCV \aket{1_k}\abra{1_k} \right)\UnitaryInteraction^\dagger\bigr] \aket{\Probe}\abra{\Probe}\Pauli_x \nonumber\\
	%&+ \trace\bigl[\UnitaryInteraction\left(\StateCR \otimes \aket{1_k}\abra{1_k} \StateCV \aket{0_k}\abra{0_k} \right)\UnitaryInteraction^\dagger\bigr] \Pauli_x\aket{\Probe}\abra{\Probe} \nonumber\\
	%&+ \trace\bigl[\UnitaryInteraction\left(\StateCR \otimes \aket{1_k}\abra{1_k} \StateCV \aket{1_k}\abra{1_k} \right)\UnitaryInteraction^\dagger\bigr] \Pauli_x\aket{\Probe}\abra{\Probe}\Pauli_x \nonumber\\
	=\; & \sum_{n,m=0}^{1} \trace\bigl[\UnitaryInteraction\bigl(\StateCR \otimes \ket{n_k}\bra{n_k}  \StateCV  \ket{m_k}\bra{m_k} \bigr)\UnitaryInteraction^\dagger\bigr] \Pauli^n_x  \ket{\Probe}\bra{\Probe}  \Pauli_x^{m}.
\end{align}
Computing the expectation values {\via} (\ref{eq:expectation}) with (\ref{eq:connection}) yields
\begin{align}
	\Expectation_k &= \bra{0_k}  \StateCV  \ket{0_k} - \bra{1_k}  \StateCV  \ket{1_k} \nonumber\\
	&= \trace\left[\Pauli_k\StateCV\right], \label{eq:expectation_D-CTCs}
\end{align}
which of course exactly reconstructs the CV state $\StateCV$. This means that focus question (ii) can be answered in the affirmative. Note that, in line with the theory, depending on the interaction $\UnitaryInteraction$, this result may not be unique, as per the uniqueness ambiguity regarding fixed points of the D-CTC CV map (\ref{eq:D-CTCs_CV}).

\subsection{Weak-measurement tomography of the state on the P-CTC}\label{sec:tomography_P-CTCs}

We now turn our attention to reconstructing the state on the P-CTC {\via} weak measurements. Our aim here is to show that one is in general able to always compute a valid P-CTC CV state for every interaction $\UnitaryInteraction$ and input $\StateCR$. It is important to note that, since the two modes of a maximally entangled state $\ket{\Bell}$ (such as on the CV subsystem) are related by transposition, {\ie}, $(A \otimes \Identity)\ket{\Bell} = (\Identity \otimes A^\transpose)\ket{\Bell}$ for $A$ a linear operator, then performing tomography on the upper CV mode is sufficient to infer the state for both modes (since they are related \emph{uniquely} by transposition). Additionally, this means that, like in the D-CTCs case, we can (without loss of generality) place the probe before the interaction (as opposed to after). This is because the transposition relation (involving both the preselected and postselected maximally entangled states) indicates that the reduced states, on either the upper or lower CV mode, are the same both before and after the interaction.

\begin{figure}[h]
	\includegraphics[scale=1]{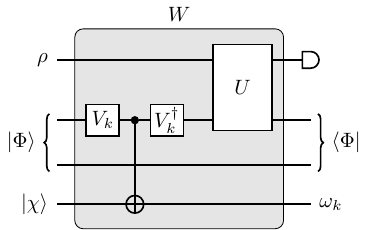}
	\caption[Weak-measurement tomography circuit for D-CTCs]{\label{fig:circuit_tomography_P-CTCs}A circuit of our tomography scheme ({\fig} \ref{fig:circuit_tomography}) incorporated in the P-CTC circuit ({\fig} \ref{fig:circuit_P-CTCs}).}
\end{figure}

Referring to the circuit diagram of the weak measurement of a P-CTC in {\fig} \ref{fig:circuit_tomography_P-CTCs}, we may express the total unitary as
\begin{equation}
	\UnitaryTotal = \bigl(\UnitaryInteraction \otimes \Identity \otimes \Identity\bigr) \cdot \bigl(\Identity \otimes \UnitaryBasis^{\dagger}_k \otimes \Identity \otimes \Identity\bigr) \cdot \CNOT{2}{4} \cdot \bigl(\Identity \otimes \UnitaryBasis_k \otimes \Identity \otimes \Identity\bigr).
\end{equation}
Evolving the composite input state $\StateCR \otimes \ket{\Bell}\bra{\Bell} \otimes \ket{\Probe}\bra{\Probe}$, subsequently postselecting the CV modes against $\Identity \otimes \ket{\Bell} \otimes \Identity$, and expanding the maximally entangled state using (without loss of generality) the particular Bell state
\begin{equation}
	\ket{\Bell} \equiv \frac{1}{\sqrt{2}}\bigl(\ket{0}\otimes\ket{0} + \ket{1}\otimes\ket{1}\bigr), \label{eq:Bell}
\end{equation}
then yields
\begin{widetext}
\begin{align}
	\StateCR \otimes \ket{\Probe}\bra{\Probe} \rightarrow \;& \bigl(\Identity \otimes \bra{\Bell} \otimes \Identity\bigr)\UnitaryTotal\bigl(\StateCR \otimes \ket{\Bell}\bra{\Bell} \otimes \ket{\Probe}\bra{\Probe}\bigr)\UnitaryTotal^\dagger\bigl(\Identity \otimes \ket{\Bell} \otimes \Identity\bigr) \nonumber\\
	%=\; & \frac{1}{4}\biggl\{\trace_2\bigl[\UnitaryInteraction\left(\Identity\otimes\aket{0_k}\abra{0_k}\right)\bigr]\StateCR\trace_2\bigl[\left(\Identity\otimes\aket{0_k}\abra{0_k}\right)\UnitaryInteraction^\dagger\bigr] \otimes \aket{\Probe}\abra{\Probe} \nonumber\\
	%&+ \trace_2\bigl[\UnitaryInteraction\left(\Identity\otimes\aket{0_k}\abra{0_k}\right)\bigr]\StateCR\trace_2\bigl[\left(\Identity\otimes\aket{1_k}\abra{1_k}\right)\UnitaryInteraction^\dagger\bigr] \otimes \aket{\Probe}\abra{\Probe}\Pauli_x \nonumber\\
	%&+ \trace_2\bigl[\UnitaryInteraction\left(\Identity\otimes\aket{1_k}\abra{1_k}\right)\bigr]\StateCR\trace_2\bigl[\left(\Identity\otimes\aket{0_k}\abra{0_k}\right)\UnitaryInteraction^\dagger\bigr] \otimes \Pauli_x \aket{\Probe}\abra{\Probe} \nonumber\\
	%&+ \trace_2\bigl[\UnitaryInteraction\left(\Identity\otimes\aket{1_k}\abra{1_k}\right)\bigr]\StateCR\trace_2\bigl[\left(\Identity\otimes\aket{1_k}\abra{1_k}\right)\UnitaryInteraction^\dagger\bigr] \otimes \Pauli_x\aket{\Probe}\abra{\Probe}\Pauli_x \biggr\} \nonumber\\
	= \; & \frac{1}{4}\sum_{n,m=0}^{1} \trace_2\bigl[\UnitaryInteraction\bigl(\Identity\otimes\ket{n_k}\bra{n_k}\bigr)\bigr]  \StateCR\,\trace_2\bigl[\bigl(\Identity\otimes\ket{m_k}\bra{m_k}\bigr)\UnitaryInteraction^\dagger\bigr] \otimes \Pauli^n_x  \ket{\Probe}\bra{\Probe}  \Pauli_x^{m}.
\end{align}
Here, the coefficient $\frac{1}{4}$ is exactly the factor that vanishes due to the prescription's renormalisation, and so for brevity we shall preemptively drop it. Accordingly, tracing out the probe yields the evolved CR state,
\begin{align}
	\StateCR_{\MapPCTCsCR} =\; & \trace_2\bigl[\bigl(\Identity \otimes \bra{\Bell} \otimes \Identity\bigr)\UnitaryTotal\bigl(\StateCR \otimes \ket{\Bell}\bra{\Bell} \otimes \ket{\Probe}\bra{\Probe}\bigr)\UnitaryTotal^\dagger\bigl(\Identity \otimes \ket{\Bell} \otimes \Identity\bigr)\bigr] \nonumber\\
	%=\; & \left(\trace_2\bigl[\UnitaryInteraction\left(\Identity\otimes\aket{0_k}\abra{0_k}\right)\bigr] + \trace_2\bigl[\UnitaryInteraction\left(\Identity\otimes\aket{1_k}\abra{1_k}\right)\bigr]\right)\StateCR\left(\trace_2\bigl[\left(\Identity\otimes\aket{0_k}\abra{0_k}\right)\UnitaryInteraction^\dagger\bigr] + \trace_2\bigl[\left(\Identity\otimes\aket{1_k}\abra{1_k}\right)\UnitaryInteraction^\dagger\bigr]\right) \nonumber\\
	%& - \left(\frac{\Strength^2}{2} - \mathcal{O}(\Strength^4)\right)\left\{\trace_2\bigl[\UnitaryInteraction\left(\Identity\otimes\aket{0_k}\abra{0_k}\right)\bigr]\StateCR\trace_2\bigl[\left(\Identity\otimes\aket{1_k}\abra{1_k}\right)\UnitaryInteraction^\dagger\bigr] + \trace_2\bigl[\UnitaryInteraction\left(\Identity\otimes\aket{1_k}\abra{1_k}\right)\bigr]\StateCR\trace_2\bigl[\left(\Identity\otimes\aket{0_k}\abra{0_k}\right)\UnitaryInteraction^\dagger\bigr]\right\} \nonumber\\
	\propto\; & \trace_\CV[\UnitaryInteraction]\, \StateCR \, \trace_\CV[\UnitaryInteraction^\dagger] + \mathcal{O}(\Strength^2).
\end{align}
We note that for small $\Strength$, the P-CTC CR evolution reduces to the usual result (\ref{eq:P-CTCs_CR}), which indicates that, to first order in the measurement strength, the P-CTC is not perturbed by the presence of a weakly interacting probe, exactly like in the D-CTCs case.

Alternatively, performing a partial trace on the CR mode yields the evolved probe state,
\begin{align}
	\StateProbe_k =\; & \trace_1\bigl[\bigl(\Identity \otimes \bra{\Bell} \otimes \Identity\bigr)\UnitaryTotal\bigl(\StateCR \otimes \ket{\Bell}\bra{\Bell} \otimes \ket{\Probe}\bra{\Probe}\bigr)\UnitaryTotal^\dagger\bigl(\Identity \otimes \ket{\Bell} \otimes \Identity\bigr)\bigr] \nonumber\\
	%=\; & \trace\left\{\trace_2\bigl[\UnitaryInteraction\left(\Identity\otimes\aket{0_k}\abra{0_k}\right)\bigr]\StateCR\trace_2\bigl[\left(\Identity\otimes\aket{0_k}\abra{0_k}\right)\UnitaryInteraction^\dagger\bigr]\right\} \aket{\Probe}\abra{\Probe} \nonumber\\
	%&+ \trace\left\{\trace_2\bigl[\UnitaryInteraction\left(\Identity\otimes\aket{0_k}\abra{0_k}\right)\bigr]\StateCR\trace_2\bigl[\left(\Identity\otimes\aket{1_k}\abra{1_k}\right)\UnitaryInteraction^\dagger\bigr]\right\} \aket{\Probe}\abra{\Probe}\Pauli_x \nonumber\\
	%&+ \trace\left\{\trace_2\bigl[\UnitaryInteraction\left(\Identity\otimes\aket{1_k}\abra{1_k}\right)\bigr]\StateCR\trace_2\bigl[\left(\Identity\otimes\aket{0_k}\abra{0_k}\right)\UnitaryInteraction^\dagger\bigr]\right\} \Pauli_x \aket{\Probe}\abra{\Probe} \nonumber\\
	%&+ \trace\left\{\trace_2\bigl[\UnitaryInteraction\left(\Identity\otimes\aket{1_k}\abra{1_k}\right)\bigr]\StateCR\trace_2\bigl[\left(\Identity\otimes\aket{1_k}\abra{1_k}\right)\UnitaryInteraction^\dagger\bigr]\right\} \Pauli_x\aket{\Probe}\abra{\Probe}\Pauli_x \nonumber\\
	=\; & \sum_{n,m=0}^{1} \trace\Bigl\{\trace_2\bigl[\UnitaryInteraction\bigl(\Identity\otimes\ket{n_k}\bra{n_k}\bigr)\bigr]\StateCR\,\trace_2\bigl[\bigl(\Identity\otimes\ket{m_k}\bra{m_k}\bigr)\UnitaryInteraction^\dagger\bigr]\Bigr\} \Pauli^n_x  \ket{\Probe}\bra{\Probe}  \Pauli_x^{m},
\end{align}
which, {\via} (\ref{eq:expectation}), corresponds to the expectation values
\begin{align}
	\Expectation_k &= \trace\Bigl\{\trace_2\bigl[\UnitaryInteraction\bigl(\Identity\otimes\ket{0_k}\bra{0_k}\bigr)\bigr[\StateCR\,\trace_2\bigl[\bigl(\Identity\otimes\ket{0_k}\bra{0_k}\bigr)\UnitaryInteraction^\dagger\bigr[\Bigr\} \nonumber\\
	&\quad - \trace\Bigl\{\trace_2\bigl[\UnitaryInteraction\bigl(\Identity\otimes\ket{1_k}\bra{1_k}\bigr)\bigr[\StateCR\,\trace_2\bigl[\bigl(\Identity\otimes\ket{1_k}\bra{1_k}\bigr)\UnitaryInteraction^\dagger\bigr[\Bigr\} \nonumber\\
	&= \sum_{n=0}^{1} (-1)^n\trace\Bigl\{\trace_2\bigl[\UnitaryInteraction\bigl(\Identity\otimes\ket{n_k}\bra{n_k}\bigr)\bigr[\StateCR\,\trace_2\bigl[\bigl(\Identity\otimes\ket{n_k}\bra{n_k}\bigr)\UnitaryInteraction^\dagger\bigr[\Bigr\}, \quad k=1,2,3. \label{eq:expectation_P-CTCs}
\end{align}
\end{widetext}
Our goal is to cast this into a form reminiscent of the right-hand side of the expression (\ref{eq:connection}), thereby allowing us to construct the trapped state in terms of the density matrix next to the $\Pauli_k$ inside the trace. Accomplishing this is a fairly involved calculation, the details of which are in {\app} \ref{app:simplification}. One obtains
\begin{equation}
	\Expectation_k = \trace[\Pauli_k \tilde{\StateCV}] \label{eq:expectation_P-CTCs_simplified}
\end{equation}
where
\begin{equation}
	\tilde{\StateCV} \equiv \frac{1}{2}(\StateIntermediate + \StateIntermediate^\dagger) \label{eq:tomography_P-CTC}
\end{equation}
with
\begin{equation}
	\StateIntermediate \equiv \trace_{1,2}\bigl[\UnitaryInteraction^{1,2} \StateCR^{1} {\UnitaryInteraction^{\dagger{1,3}}}\bigr].
\end{equation}
Here, the superscripts indicate the subsystems upon which each operator acts. As per the reconstruction scheme specified in {\secn}(\ref{sec:tomography}), the state $\tilde{\StateCV}$ may be written as
\begin{align}
	\tilde{\StateCV} &= \frac{1}{2} \sum_{k=0}^{3} \trace[\Pauli_k \tilde{\StateCV}]\Pauli_k \nonumber\\
	&= \frac{1}{2}\biggl(\trace[\tilde{\StateCV}]\Identity + \sum_{k=1}^{3} \trace[\Pauli_k \tilde{\StateCV}]\Pauli_k\biggr).
\end{align}
Using this and the expectation values (\ref{eq:expectation_P-CTCs_simplified}) allows us to reconstruct the trapped state $\StateCV_{\MapPCTCsCR}^{(\mathrm{w})}$ in a similar fasion,
\begin{align}
	\StateCV_{\MapPCTCsCR}^{(\mathrm{w})} &= \frac{1}{2}\biggl(\Identity + \sum_{k=1}^{3} \Expectation_k \Pauli_k\biggr) \nonumber\\
	&= \frac{1}{2}\left(1 - \trace[\tilde{\StateCV}]\right)\Identity + \tilde{\StateCV}. \label{eq:P-CTCs_CV_tomography}
\end{align}
where, by requirement for physicality, we assumed that $\StateCV_{\MapPCTCsCR}^{(\mathrm{w})}$ has unity trace, {\ie}, $\trace\bigl[\StateCV_{\MapPCTCsCR}^{(\mathrm{w})}\bigr] = 1$.

Thus, our result here specifies the state (\ref{eq:P-CTCs_CV_tomography}) to be the one which successive weak measurements of the state on the P-CTC would reveal. This therefore answers our third focus question (iii) in the affirmative.

\subsection{Na\"{i}ve computation of the state on the P-CTC}\label{sec:P-CTCs_CV}

In this section, we explore whether the P-CTC state can be determined entirely from simple analysis of the prescription itself, rather than from elaborate methods like weak-measurement tomography from the preceding section. Rather surprisingly, this turns out to be entirely possible. We therefore present here the simple construction of the state, and subsequently prove it to be equivalent to the result from our weak-measurement tomography ({\secn} \ref{sec:tomography_P-CTCs}).

\begin{figure}[h]
	\includegraphics[scale=1]{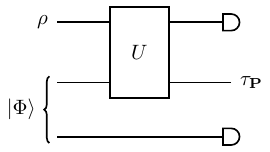}
	\caption[P-CTC chronology-violating state]{\label{fig:circuit_P-CTCs_CV}A circuit diagram depicted our methodology for isolating the P-CTC CV state $\StateCV_{\MapPCTCsCR}$. The rounded caps on the uppermost and lowermost modes represent partial traces on those modes.}
\end{figure}

Figure \ref{fig:circuit_P-CTCs_CV} depicts the following analysis. Beginning with the evolved state spanning the CR and CV state in the P-CTCs prescription (\ref{eq:P-CTCs_evolved}), we first discard the CR mode by using the partial trace operation, which yields the evolution of the maximally entangled state on the CV and teleportation modes,
\begin{equation}
	\ket{\Bell}\bra{\Bell} \rightarrow \trace_\CR\bigl[\bigl(\UnitaryInteraction \otimes \Identity\bigr)\bigl(\StateCR \otimes \ket{\Bell}\bra{\Bell}\bigr)\bigl(\UnitaryInteraction^\dagger \otimes \Identity\bigr)\bigr].
\end{equation}
After throwing out the teleportation subsystem, the post-interaction CV state is then obtained,
\begin{align}
	\StateCV_{\MapPCTCsCR} &= \trace_\mathrm{CR,T}\bigl[\bigl(\UnitaryInteraction \otimes \Identity\bigr)\bigl(\StateCR \otimes \ket{\Bell}\bra{\Bell}\bigr)\bigl(\UnitaryInteraction^\dagger \otimes \Identity\bigr)\bigr] \nonumber\\
	&= \trace_\CR\bigl[\UnitaryInteraction(\StateCR \otimes \tfrac{1}{2}\Identity)\UnitaryInteraction^\dagger\bigr]. \label{eq:P-CTCs_CV}
\end{align}
Comparing this with the D-CTCs self-consistency condition (\ref{eq:D-CTCs_CV}), we see that it may be re-expressed in terms of the D-CTC CV map (\ref{eq:D-CTCs_CV}), yielding
\begin{equation}
	\StateCV_{\MapPCTCsCR} = \MapDCTCsCV_{U}\left[\StateCR,\tfrac{1}{2}\Identity\right].
\end{equation}
This is a curious result as, according to the ECP ({\secn} \ref{sec:D-CTCs}), it suggests that the P-CTC CV state is simply the first D-CTC iteration of the maximally mixed seed state $\StateCV^{(0)} = \frac{1}{2}\Identity$.

To demonstrate that this `na\"{i}ve' P-CTC state (\ref{eq:P-CTCs_CV}) is equivalent to our tomographically inferred state (\ref{eq:P-CTCs_CV_tomography}), we simply show how the expectation values of the former,
\begin{equation}
	\Expectation_k = \trace[\Pauli_k \StateCV_{\MapPCTCsCR}] \label{eq:expectation_P-CTCs_CV}
\end{equation}
are equivalent to those for $\StateCV_{\MapPCTCsCR}^{(\mathrm{w})}$, that is,
\begin{equation}
	\trace[\Pauli_k \tilde{\StateCV}] = \trace[\Pauli_k \StateCV_{\MapPCTCsCR}].
\end{equation}
This calculation appears in {\app} \ref{app:equivalency}. Since $\StateCV_{\MapPCTCsCR}$ is normalised, then (\ref{eq:expectation_P-CTCs_CV}) will reconstruct the same state as (\ref{eq:expectation_P-CTCs_simplified}), and so the state on the P-CTC must have exactly the form
\begin{equation}
	\StateCV_{\MapPCTCsCR} = \frac{1}{2}\trace_{1}\bigl[\UnitaryInteraction^{1,2} \StateCR^{1} {\UnitaryInteraction^{\dagger{1,2}}}\bigr],
\end{equation}
{\ie}, {\eqn} (\ref{eq:P-CTCs_CV}). Note that this exists for any given $\StateCR$ and $\UnitaryInteraction$, and provides a single solution for every pairing. %However, we restate that this (tomography) result only works for qubits, as an equivalent expression for qudits seems to be difficult to determine (see {\secn} \ref{sec:conclusion}).

We therefore conclude that our tomography scheme for P-CTCs recovers an operational CV state that is exactly equivalent to the state determined from a very na\"{i}ve analysis of P-CTCs (in which all modes but the CV mode are traced away). The fact that it takes the form of the ECP first iteration state (with maximally mixed seed) suggests that it obeys a similar self-consistency condition to D-CTCs, which hints at a partial answer for focus question (iv) (more discussion on this in {\secn} \ref{sec:P-CTC_fixed_points}).

\section{Examples}\label{sec:results}

In this section, we apply our findings to the specific cases of three archetypal interactions. Of these, the grandfather paradox provides a simple example of a nontrivial interaction in which the P-CTC and D-CTC CV states are equivalent, while the unproven theorem paradox allows us to analyse a circuit in which parametrised D-CTC solutions arise. Lastly, the probabilistic \textgate{swap} interaction presents a useful example of a scenario in which the P-CTC solution is not a D-CTC fixed point in general.

\subsection{Grandfather paradox}\label{sec:grandfather}

\begin{figure}[h]
	\includegraphics[scale=1]{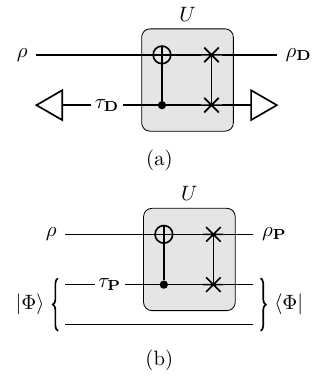}
	\caption[Grandfather paradox circuits]{\label{fig:circuit_grandfather}Circuit diagrams of the grandfather paradox in the (a) D-CTC and (b) P-CTC prescriptions.}
\end{figure}

Probably the most prominent time-travel paradox is the grandfather paradox. We present here a simplified version (based on \cite{lloyd_closed_2011}) in which a qubit CV state has an apparently paradoxical action on its past self (exterior to the CTC) {\via} a controlled-\textgate{not} operation. Illustrated in {\fig} \ref{fig:circuit_grandfather}, the corresponding unitary form is
\begin{equation}
	\UnitaryInteraction = \Swap \cdot \CNOT{2}{1}. \label{eq:unitary_grandfather}
\end{equation}
where
\begin{equation}
	\Swap \equiv \sum_{i,j=0}^{1}\ket{i}\bra{j}\otimes\ket{j}\bra{i}.
\end{equation}
is the \textgate{swap} gate. It is straightforward to calculate both pairs of D-CTC and P-CTC states,
\begin{subequations}
\begin{align}
	\StateCR_{\MapDCTCsCR} &= \frac{1}{2}\Bigl[\ket{0}\bra{0} + \bigl(\bra{0}\StateCR\ket{1} + \bra{1}\StateCR\ket{0}\bigr)^2\bigl(\ket{0}\bra{1} + \ket{1}\bra{0}\bigr) + \ket{1}\bra{1}\Bigr], \label{eq:D-CTCs_CR_grandfather} \\
	\StateCV_{\MapDCTCsCR} &= \frac{1}{2}\StateCR + \frac{1}{2}\Pauli_x\StateCR\Pauli_x; \label{eq:D-CTCs_CV_grandfather}
\end{align}
\end{subequations}
\vspace{-0.75cm}
\begin{subequations}
\begin{align}
	\StateCR_{\MapPCTCsCR} &= \ket{+}\bra{+}, \label{eq:P-CTCs_CR_grandfather} \\
	\StateCV_{\MapPCTCsCR} &= \frac{1}{2}\StateCR + \frac{1}{2}\Pauli_x\StateCR\Pauli_x. \label{eq:P-CTCs_CV_grandfather}
\end{align}
\end{subequations}
Immediately, we can see that, while the CR output states are distinct, the CV states are the same, which means that the P-CTC CV state is the unique fixed point of the D-CTC self-consistency map for all $\StateCR$. Note however that here the P-CTC CR state (\ref{eq:P-CTCs_CR_grandfather}) only exists if $\bra{0}\StateCR\ket{0} \neq 0$ since the P-CTC operator from (\ref{eq:P-CTCs_CR}) in this case takes the nonunitary form $\trace_{\CV}[\Unitary] = \ket{+}\bra{0}$. Thus, if $\bra{0}\StateCR\ket{0} = 0$, the P-CTC prescription fails to provide a solution to the paradox (yet our methodology is still able to compute a P-CTC CV state).

\subsection{Unproven theorem paradox}\label{sec:unproven}

\begin{figure}[h]
	\includegraphics[scale=1]{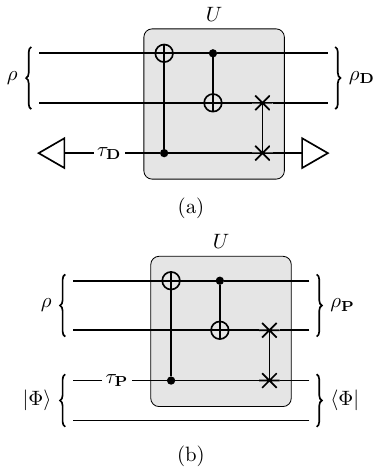}
	\caption[Unproven theorem paradox circuits]{\label{fig:circuit_unproven}Circuit diagrams of the unproven theorem paradox in the (a) D-CTC and (b) P-CTC prescriptions.}
\end{figure}

The unproven theorem paradox \cite{deutsch_quantum_1991, lloyd_closed_2011, allen_treating_2014} (depicted in {\fig} \ref{fig:circuit_unproven}) is, as we shall see, not as straightforward as the previous interaction. Here, the CR subsystem is comprised of two modes: the first is the `book', and the second is the `mathematician'. Described by the unitary
\begin{equation}
	\UnitaryInteraction = \Swap^{2,3} \cdot \CNOT{1}{2} \cdot \CNOT{3}{1}, \label{eq:unitary_unproven}
\end{equation}
the unproven theorem paradox encapsulates the scenario in which the proof of a theorem that a mathematician reads in the past is the very same proof that they write down in the future book (with the book having travelled back in time). We therefore denote the absence and presence of the proof in each subsystem intuitively by $\ket{0}$ and $\ket{1}$ respectively. Under this interpretation, when the mathematician and/or the book are in the state $\ket{1}$, they possess a `correct' proof, while $\ket{0}$ means that they do not. (Alternatively, one can think of these qubit levels as denoting different answers to some problem the mathematician is working on. They can then encode their proof of some theorem into an $\NumberLevels$-bit binary string by using $\NumberLevels$ copies of the circuit \cite{allen_treating_2014}.) Thus, with the CR input state
\begin{equation}
	\StateCR = \ket{0}\bra{0}\otimes\ket{0}\bra{0}
\end{equation}
({\ie}, both book and mathematician do not initially possess the proof), the D-CTC and P-CTC solutions are
\begin{subequations}
\begin{align}
	\StateCR_{\MapDCTCsCR} &= \ParameterFree\ket{0}\bra{0}\otimes\ket{0}\bra{0} + (1-\ParameterFree)\ket{1}\bra{1}\otimes\ket{1}\bra{1}, \label{eq:D-CTCs_CR_unproven}\\
	\StateCV_{\MapDCTCsCR} &= \ParameterFree\ket{0}\bra{0} + (1-\ParameterFree)\ket{1}\bra{1}, \quad 0 \leq \ParameterFree \leq 1; \label{eq:D-CTCs_CV_unproven}
\end{align}
\end{subequations}
\vspace{-0.75cm}
\begin{subequations}
\begin{align}
	\StateCR_{\MapPCTCsCR} &= \ket{\Bell^+}\bra{\Bell^+}, \label{eq:P-CTCs_CR_unproven}\\
	\StateCV_{\MapPCTCsCR} &= \frac{1}{2}\Identity. \label{eq:P-CTCs_CV_unproven}
\end{align}
\end{subequations}
Observe that the D-CTC solution turns out to be nonunique (with free parameter $\ParameterFree$), while the P-CTC CV state is not. It is easy to see however that the latter is the maximally entropic form of its D-CTC counterpart, which means that the ECP end-state evolution (the `correct' D-CTC fixed point which resolves the uniqueness ambiguity) in this case coincidentally is exactly the P-CTC solution (\ref{eq:P-CTCs_CV_unproven}).

\subsection{Probabilistic scattering}\label{sec:swap}

The last example which we will explore is that of a basic probabilistic scattering interaction. Our specific analysis will concern the power-of-\textgate{swap} gate \cite{williams_explorations_2010, bishop_billiard-ball_2022},
\begin{equation}
	\Swap^\ParameterPower = \frac{1 + \e^{-\eye\pi \ParameterPower}}{2}\Identity\otimes\Identity + \frac{1 - \e^{-\eye\pi \ParameterPower}}{2}\Swap, \quad \ParameterPower \in \Reals \label{eq:unitary_swap}
\end{equation}
where the parameter $\ParameterPower$ represents the finite probability for a pair of interacting particles to collide with each other. This is because the action of this form on a pure bipartite composition $\ket{\psi_\mathrm{A}}\otimes\ket{\psi_\mathrm{B}}$ is
\begin{align}
	\Swap^\ParameterPower \ket{\psi_\mathrm{A}}\otimes\ket{\psi_\mathrm{B}} &= \frac{1 + \e^{-\eye\pi \ParameterPower}}{2}\ket{\psi_\mathrm{A}}\otimes\ket{\psi_\mathrm{B}} \nonumber\\
	&\quad + \frac{1 - \e^{-\eye\pi \ParameterPower}}{2}\ket{\psi_\mathrm{B}}\otimes\ket{\psi_\mathrm{A}},
\end{align}
which is interpretable as the spontaneous, probabilistic diffusion of a pair of systems between two modes. Thus, despite not being a temporal paradox in the typical sense, the probabilistic scattering interaction makes for an interesting scenario as it perhaps has a more realistic physical interpretation than the previously studied interactions. Furthermore, as we will see, it provides an example where the CV states of D-CTCs and P-CTCs do not coincide.

\begin{figure}[b]
	\includegraphics[scale=1]{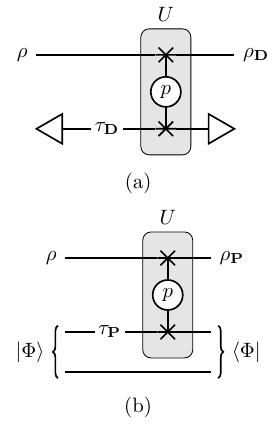}
	\caption[Probabilistic swap circuits]{\label{fig:circuit_swap}Circuit diagrams of the \textgate{swap} interaction in the (a) D-CTC and (b) P-CTC prescriptions.}
\end{figure}

In contrast with the relatively straightforward computations of the other examples, the D-CTC analysis for our probabilistic \textgate{swap} interaction here (illustrated in {\fig} \ref{fig:circuit_swap}) is slightly more involved. With (\ref{eq:unitary_swap}) as our unitary interaction, {\ie},
\begin{equation}
	\UnitaryInteraction = \Swap^\ParameterPower,
\end{equation}
we can compute the evolution of the input states as
\begin{align}
	\Swap^\ParameterPower(\StateCR \otimes \StateCV_{\MapDCTCsCR})\Swap^{\ParameterPower \dagger} &= \cos^2\left(\frac{\pi \ParameterPower}{2}\right)\StateCR \otimes \StateCV_{\MapDCTCsCR} \nonumber\\
	&\quad -\frac{\eye}{2}\sin(\pi \ParameterPower)\left(\StateCR\otimes\StateCV_{\MapDCTCsCR}\right)\Swap \nonumber\\
	&\quad +\frac{\eye}{2}\sin(\pi \ParameterPower)\Swap\left(\StateCR\otimes\StateCV_{\MapDCTCsCR}\right) \nonumber\\
	&\quad + \sin^2\left(\frac{\pi \ParameterPower}{2}\right)\StateCV_{\MapDCTCsCR} \otimes \StateCR.
\end{align}
From this, we use the identities
\begin{subequations}
\begin{align}
	\trace_\CR\bigl[\left(\StateCR\otimes\StateCV_{\MapDCTCsCR}\right)\Swap\bigr] &= \StateCV_{\MapDCTCsCR}\StateCR, \\
	\trace_\CR\bigl[\Swap\left(\StateCR\otimes\StateCV_{\MapDCTCsCR}\right)\bigr] &= \StateCR\StateCV_{\MapDCTCsCR},
\end{align}
\end{subequations}
to compute the D-CTC fixed point condition
\begin{align}
	\StateCV_{\MapDCTCsCR} &= \trace_\CR\bigl[\Swap^\ParameterPower\left(\StateCR\otimes\StateCV_{\MapDCTCsCR}\right)\Swap^{\ParameterPower\dagger}\bigr] \nonumber\\
	&= \StateCR - \eye \cot\left(\frac{\pi \ParameterPower}{2}\right)\left[\StateCV_{\MapDCTCsCR},\StateCR\right] \label{eq:consistency_swap}
\end{align}
where $\cot({}\cdot{})$ is the cotangent function, and
\begin{equation}
	\left[A,B\right] = AB - BA
\end{equation}
is the commutator. Prior to this computation, it is clear that for even-integer $\ParameterPower$, the corresponding interaction $\UnitaryInteraction = \Identity$ is trivial. In this case, no restrictions are placed on the trapped state $\StateCV_{\MapDCTCsCR}$ (other than requiring $\trace[\StateCV_{\MapDCTCsCR}] = 1$), and the output state is simply $\StateCR_{\MapDCTCsCR} = \StateCR$. Conversely, when $\ParameterPower$ is not an even integer, computational symbolic algebra solvers find $\StateCV_{\MapDCTCsCR} = \StateCR$ to be the unique solution of {\eqn} (\ref{eq:consistency_swap}). The P-CTC states may be similarly computed, and to summarise,
\begin{subequations}
\begin{align}
	\StateCR_{\MapDCTCsCR} &= \StateCR, \label{eq:D-CTCs_CR_swap}\\
	\StateCV_{\MapDCTCsCR} &= \StateCR, \quad (\ParameterPower \neq 2k, k\in\Integers); \label{eq:D-CTCs_CV_swap}
\end{align}
\end{subequations}
\vspace{-0.75cm}
\begin{subequations}
\begin{align}
	\StateCR_{\MapPCTCsCR} &= \StateCR, \label{eq:P-CTCs_CR_swap}\\
	\StateCV_{\MapPCTCsCR} &= \cos^2\left(\frac{\pi \ParameterPower}{2}\right)\frac{\Identity}{2} + \sin^2\left(\frac{\pi \ParameterPower}{2}\right)\StateCR. \label{eq:P-CTCs_CV_swap}
\end{align}
\end{subequations}
These are discussed at length in the next section (\ref{sec:P-CTC_fixed_points}).

\section{Discussion}\label{sec:discussion}

%\subsection{P-CTC state as a D-CTC fixed point}\label{sec:P-CTC_fixed_points}
\label{sec:P-CTC_fixed_points}

From our P-CTC CV analysis in {\secn} \ref{sec:P-CTCs_CV} and subsequent weak tomography in {\secn} \ref{sec:tomography_P-CTCs}, we can conclude that weak measurements recover a P-CTC CV state (\ref{eq:P-CTCs_CV}) that may be interpreted as the first iteration of maximally mixed state in the ECP ({\secn} \ref{sec:D-CTCs}). In essence, this means that a P-CTC can be thought of as being somewhat like a `partial' D-CTC {\textendash} one in which the self-consistency condition is not fully realised (at least according to D-CTCs). Here, we discuss the similarities and differences between the CV states of the two prescriptions, thereby answering our fourth (iv) focus question from {\secn} \ref{sec:introduction}.

Perhaps the most obvious place to begin is to determine whether the state on the P-CTC is a fixed point of the D-CTC CV map (\ref{eq:D-CTCs_CV_fixed}). According to our above analysis, this is true only when the D-CTC CV state converges to a fixed point in the first iteration in the ECP. Although this turns out to be identically true for a wide variety of interactions, including the grandfather paradox (see {\secn} \ref{sec:grandfather}), it is only partially true for the unproven theorem paradox ({\secn} \ref{sec:unproven}), and not true at all in the case of the \textgate{swap} interaction of {\secn} \ref{sec:swap} when $\ParameterPower \neq 2k, k\in\Integers$ ({\ie}, non-even-integer $\ParameterPower$).

Let us consider the probabilistic \textgate{swap} interaction in more detail. Here, one concludes that $\StateCV_{\MapDCTCsCR} = \StateCR$ as per (\ref{eq:D-CTCs_CV_swap}) for all non-even-integer values of the power parameter $\ParameterPower$. This contrasts with the first iteration in the ECP with the seed state $\StateCV^{(0)} = \frac{1}{2}\Identity$, a simple computation of which reveals
\begin{equation}
	\StateCV^{(1)} = \MapDCTCsCV_{\Swap^\ParameterPower}\bigl[\StateCR,\StateCV^{(0)}\bigr] = \cos^2\left(\frac{\pi \ParameterPower}{2}\right)\StateCV^{(0)} + \sin^2\left(\frac{\pi \ParameterPower}{2}\right)\StateCR,
\end{equation}
in line with (\ref{eq:P-CTCs_CV_swap}). By using this in conjunction with the recursion identity (\ref{eq:ECP_recursion}), we can calculate the $(n+1)$th iteration
\begin{align}
	\StateCV^{(n+1)} &= \MapDCTCsCV_{\Swap^\ParameterPower}\bigl[\StateCR,\StateCV^{(n)}\bigr] \nonumber\\
	&= \cos^2\left(\frac{\pi \ParameterPower}{2}\right)\StateCV^{(n)} + \sin^2\left(\frac{\pi \ParameterPower}{2}\right)\StateCR \nonumber\\
	&= \MapDCTCsCV^{n+1}_{\Swap^\ParameterPower}\bigl[\StateCR,\StateCV^{(0)}\bigr] \nonumber\\
	&= \cos^{2(n+1)}\left(\frac{\pi \ParameterPower}{2}\right)\bigl(\StateCV^{(0)} - \StateCR\bigr) + \StateCR, \label{eq:recursion_swap}
\end{align}
where we used the series identity
\begin{equation}
	\sum_{k=0}^{n} \cos^{2k}\left(\frac{\pi \ParameterPower}{2}\right) = \frac{1 - \cos^{2(n+1)}\left(\frac{\pi \ParameterPower}{2}\right)}{\sin^2\left(\frac{\pi \ParameterPower}{2}\right)}.
\end{equation}
Note also that we introduced the notation $\MapDCTCsCV^{n+1}$ to denote $n+1$ compositions of the D-CTC CV map, {\ie}, if
\begin{equation}
	\left(\MapDCTCsCV_{\UnitaryInteraction} \circ \MapDCTCsCV_{\UnitaryInteraction}\right)\bigl[\StateCR,\StateCV^{(0)}\bigr] \equiv \MapDCTCsCV_{\UnitaryInteraction}\Bigl[\StateCR,\MapDCTCsCV_{\UnitaryInteraction}\bigl[\StateCR,\StateCV^{(0)}\bigr]\Bigr]
\end{equation}
then
\begin{align}
	\MapDCTCsCV^{n+1}_{\UnitaryInteraction}\bigl[\StateCR,\StateCV^{(0)}\bigr] &= (\,\underbrace{\MapDCTCsCV_{\UnitaryInteraction} \circ \MapDCTCsCV_{\UnitaryInteraction} \circ \ldots \circ \MapDCTCsCV_{\UnitaryInteraction}}_{n+1 \text{ times}}\,)\bigl[\StateCR,\StateCV^{(0)}\bigr] \nonumber\\
	&= \MapDCTCsCV_{\UnitaryInteraction}\biggl[ \StateCR,\MapDCTCsCV_{\UnitaryInteraction}\Bigl[ \StateCR,\ldots,\MapDCTCsCV_{\UnitaryInteraction}\bigl[\StateCR,\StateCV^{(0)}\bigr] \Bigr] \biggr].
\end{align}
It is easy to see that, in the limit $n \rightarrow \infty$ of (\ref{eq:recursion_swap}), we obtain
\begin{align}
	\StateCV^{(\infty)} &\equiv \MapDCTCsCV^{\infty}_{\Swap^\ParameterPower}\bigl[\StateCR,\StateCV^{(0)}\bigr] \nonumber\\
	&= \StateCR + \lim_{n \rightarrow \infty} \cos^{2(n+1)}\left(\frac{\pi \ParameterPower}{2}\right)\bigl(\StateCV^{(0)} - \StateCR\bigr) \nonumber\\
	&= \StateCR,
\end{align}
since $\abs{\cos^2\left(\frac{\pi \ParameterPower}{2}\right)} < 1$ when $\ParameterPower \neq 2k$ with $k\in\Integers$. As this result agrees with (\ref{eq:D-CTCs_CV_swap}), we can conclude that the probabilistic \textgate{swap} interaction is a case in which an infinite number of iterations in the ECP are required to arrive at the fixed point.

Given that $\StateCR$ is arbitrary, it is then natural to wonder if there exists a class of interactions that have an associated fixed-point P-CTC CV state. Completely characterising this in general appears to be difficult, but a simple condition can be derived with a little work. We first recall that idempotency of the D-CTC CV map (\ref{eq:D-CTCs_CV}) is the condition
\begin{equation}
	\MapDCTCsCV^n_{U}[\StateCR,\StateCV] = \MapDCTCsCV_{U}[\StateCR,\StateCV], \quad \forall \; n\in\Integers_{> 1}.
\end{equation}
With respect to a P-CTC state $\StateCV_{\MapPCTCsCR}$ at the first ($n=1$) iteration, this takes the form
\begin{equation}
	\MapDCTCsCV_{U}[\StateCR,\StateCV_{\MapPCTCsCR}] = \StateCV_{\MapPCTCsCR}.
\end{equation}
Substituting in the summation reconstruction (\ref{eq:P-CTCs_CV_tomography}) into the left-hand side, and subsequently rearranging and identifying the form (\ref{eq:P-CTCs_CV}) yields
\begin{equation}
	\sum_{k=1}^{3}\Expectation_k \MapDCTCsCV_{U}[\StateCR,\Pauli_k] = 0
\end{equation}
where $\Expectation_k$ are the familiar expectation values from {\secn} \ref{sec:tomography_P-CTCs}. This is exactly the condition that $\UnitaryInteraction$ must satisfy in order for the corresponding $\StateCV_{\MapPCTCsCR}$ to be a D-CTC fixed point.

\section{Conclusion}\label{sec:conclusion}

In this work, we employed weak-measurement tomography as a means to precisely determining the form of the states on both a D-CTC and a P-CTC. Our methodology, based on coupling an ancillary `probe' state to the CV mode {\via} a controlled-\textgate{not} interaction, proved to be successful, as it enabled us to make strong inferences about the state of the system in both prescriptions. We tested our scheme by verifying that the state on the D-CTC can be operationally recovered without (significantly) disturbing it, and followed this by demonstrating how the otherwise unassignable P-CTC state may likewise be determined. This latter state, which is necessarily compatible with the particular form of temporal self-consistency as demanded by P-CTCs, was also demonstrated to be derivable from the P-CTC prescription itself.

We explored our procedure by applying it to a few specific examples of time-travel interactions, including a simplified version of the grandfather paradox, the unproven theorem paradox, and a model of probabilistic particle scattering. We computed a P-CTC state that was both physical and unique in all cases, with some having further meaning in regard to D-CTCs. In particular, while the unproven theorem paradox had a solution parametrisation under the D-CTC description, the P-CTC CV state did not, and assumed the form of the maximally entropic D-CTC state. %We further found that, for the probabilistic scattering interaction, the P-CTC state is equivalent to the output state from the first iteration of the equivalent circuit picture (ECP, see {\secn} \ref{sec:D-CTCs}) of D-CTCs, where the seed state is maximally mixed.%A general interpretation of these findings is that in the cases where bare D-CTCs possesses a uniqueness ambiguity, the P-CTC state is equivalent to the `correct' D-CTC fixed point as prescribed by Deutsch.

%In the event where P-CTCs place constraints on initial data as a function of the future, we can conclude that the P-CTC state is unaffected. This means that, although a subset of input states are rendered forbidden purely due to their potential (but ultimately unrealised) future evolutions, a valid state to the system on the P-CTC can always be assigned.%Consequently, our results do not seem to be affected by the P-CTC peculiarity of antichronological influence.

The generalisation of our scheme to $d$-level `qudit' systems would prove to be an interesting exercise, and should provide more insight into the generality of our results, in particular whether the P-CTC CV form (\ref{eq:P-CTCs_CV}) holds true in higher dimensions. We expect that such generalisation is an entirely possible task, but does seem to be nontrivial, and so is left to future investigation.

We therefore conclude that, at least in regard to time-travelling qubits, it is possible to rigorously assign a state to the system on the P-CTC. In contrast to its D-CTC counterpart, this state is unique for every given combination of interaction and input state, and was further shown to be equivalent to the ECP's first iteration of the CV state. This is curious, as it suggests that despite their striking differences, the distinct notions of self-consistency that underpin both P-CTCs and D-CTCs may be more similar than originally thought.

\begin{acknowledgments}
	This research was supported by the Australian Research Council (ARC) under the Centre of Excellence for Quantum Computation and Communication Technology (Project No.~CE170100012). F.C.~acknowledges support through an Australian Research Council Discovery Early Career Researcher Award (DE170100712) and under the Centre of Excellence for Engineered Quantum Systems (EQUS, CE170100009). The University of Queensland (UQ) acknowledges the Traditional Owners and their custodianship of the lands on which UQ operates.
\end{acknowledgments}

\bibliographystyle{apsrev4-2}
\interlinepenalty=10000
\bibliography{quantum_state_tomography_on_closed_timelike_curves_using_weak_measurements_v2}

\onecolumngrid
\newpage
\appendix
\section{\uppercase{Simplification of the P-CTC expectation value}}\label{app:simplification}

By employing superscripts to explicitly label which Hilbert subsystems each operator acts on, we can write (\ref{eq:expectation_P-CTCs}) as
\begin{align}
	\Expectation_k =\; & \sum_{n=0}^{1} (-1)^n\trace\left\{\trace_2\left[\UnitaryInteraction^{1,2}\left(\Identity^{1}\otimes\ket{n_k}\bra{n_k}^{2}\right)\right] \left(\StateCR^{1} \otimes \Identity^{2} \otimes \Identity^{3}\right) \trace_2\left[\left(\Identity^{1}\otimes\ket{n_k}\bra{n_k}^{2}\right){\UnitaryInteraction^{\dagger{1,2}}}\right]\right\}. \label{eq:expectation_P-CTCs_2}
\end{align}
Our next step is to introduce a third subsystem, using which we can `extend' the operators in (\ref{eq:expectation_P-CTCs_2}) and equivalently write
\begin{align}
	\Expectation_k %=\; & \trace\left\{\trace_2\left[\UnitaryInteraction^{1,2}\left(\Identity^{1}\otimes\aket{0_k}\abra{0_k}^{2}\otimes\Identity^{3}\right)\right]\left(\StateCR^{1}\otimes\Identity^{2}\otimes\Identity^{3}\right)\trace_3\left[\left(\Identity^{1}\otimes\Identity^{2}\otimes\aket{0_k}\abra{0_k}^{3}\right){\UnitaryInteraction^\dagger}^{1,3}\right]\right\} \nonumber\\
	%&- \trace\left\{\trace_2\left[\UnitaryInteraction^{1,2}\left(\Identity^{1}\otimes\aket{1_k}\abra{1_k}^{2}\otimes\Identity^{3}\right)\right]\left(\StateCR^{1}\otimes\Identity^{2}\otimes\Identity^{3}\right)\trace_3\left[\left(\Identity^{1}\otimes\Identity^{2}\otimes\aket{1_k}\abra{1_k}^{3}\right){\UnitaryInteraction^\dagger}^{1,3}\right]\right\} \nonumber\\
	=\; & \sum_{n=0}^{1} (-1)^n\trace\left\{\trace_2\left[\UnitaryInteraction^{1,2}\left(\Identity^{1}\otimes\ket{n_k}\bra{n_k}^{2}\otimes\Identity^{3}\right)\right] \left(\StateCR^{1} \otimes \Identity^{2} \otimes \Identity^{3}\right) \trace_3\left[\left(\Identity^{1}\otimes\Identity^{2}\otimes\ket{n_k}\bra{n_k}^{3}\right){\UnitaryInteraction^{\dagger{1,3}}}\right]\right\}. \label{eq:expectation_P-CTCs_3}
\end{align}
It is important to note that the third subsystem here (denoted by a superscript number 3), despite being a mode in the CV subsystem, is not to be misconstrued as the lower arm of the pair which forms the P-CTC. Rather, we introduce it as a mathematical device by which we can combine the partial trace operations. Essentially, because said trace operations act on different Hilbert spaces, they can be amalgamated along with the state $\StateCR^{1}$, {\eg},
\begin{align}
	\trace_2\left[\UnitaryInteraction^{1,2}\left(\Identity^{1} \otimes \ket{n_k}\bra{n_k}^{2} \otimes \Identity^{3}\right)\right] \left(\StateCR^{1} \otimes \Identity^{2} \otimes \Identity^{3}\right) \trace_3\left[\left(\Identity^{1} \otimes \Identity^{2} \otimes \ket{n_k}\bra{n_k}^{3}\right){\UnitaryInteraction^{\dagger{1,3}}}\right] \nonumber\\
	= \trace_{2,3}\left[\UnitaryInteraction^{1,2}\left(\StateCR^{1} \otimes \ket{n_k}\bra{n_k}^{2} \otimes \ket{n_k}\bra{n_k}^{3}\right){\UnitaryInteraction^{\dagger{1,3}}}\right].
\end{align}
Using this in (\ref{eq:expectation_P-CTCs_3}) therefore yields
\begin{align}
	\Expectation_k %=\; & \trace\left[\UnitaryInteraction^{1,2}\left(\StateCR^{1} \otimes \aket{0_k}\abra{0_k}^{2} \otimes \aket{0_k}\abra{0_k}^{3}\right){\UnitaryInteraction^\dagger}^{1,3} - \UnitaryInteraction^{1,2}\left(\StateCR^{1} \otimes \aket{1_k}\abra{1_k}^{2} \otimes \aket{1_k}\abra{1_k}^{3}\right){\UnitaryInteraction^\dagger}^{1,3}\right] \nonumber\\
	%=\; & \trace\left\{\trace_{2,3}\left[\UnitaryInteraction^{1,2}\left(\StateCR^{1} \otimes \aket{0_k}\abra{0_k}^{2} \otimes \aket{0_k}\abra{0_k}^{3}\right){\UnitaryInteraction^\dagger}^{1,3}\right] - \trace_{2,3}\left[\UnitaryInteraction^{1,2}\left(\StateCR^{1} \otimes \aket{1_k}\abra{1_k}^{2} \otimes \aket{1_k}\abra{1_k}^{3}\right){\UnitaryInteraction^\dagger}^{1,3}\right]\right\} \nonumber\\
	=\; & \sum_{n=0}^{1} (-1)^n\trace\left[\UnitaryInteraction^{1,2}\left(\StateCR^{1} \otimes \ket{n_k}\bra{n_k}^{2} \otimes \ket{n_k}\bra{n_k}^{3}\right){\UnitaryInteraction^{\dagger{1,3}}}\right]. \label{eq:expectation_P-CTCs_4}
\end{align}
With foresight, we then make use of the cyclic property of the trace operation, or more specifically, a weaker property of cyclicity for the partial trace, {\eg},
\begin{equation}
	\trace_{m}\left[A^{n,m} B^{m}\right] = \trace_{m}\left[B^{m} A^{n,m}\right]
\end{equation}
to shift the basis states $\ket{n_k}\bra{n_k}$ out to the front of the product inside the trace,
\begin{align}
	\Expectation_k =\; & \sum_{n=0}^{1} (-1)^n\trace\left[\left(\Identity^{1} \otimes \ket{n_k}\bra{n_k}^{2} \otimes \ket{n_k}\bra{n_k}^{3}\right) \UnitaryInteraction^{1,2} \StateCR^{1} {\UnitaryInteraction^{\dagger{1,3}}}\right]. \label{eq:expectation_P-CTCs_5}
\end{align}
Next, we use the key identity,
\begin{equation}
	\sum_{n=0}^{1} (-1)^n \ket{n_k}\bra{n_k} \otimes \ket{n_k}\bra{n_k} \iffalse= \ket{0_k}\bra{0_k} \otimes \ket{0_k}\bra{0_k} - \ket{1_k}\bra{1_k} \otimes \ket{1_k}\bra{1_k} \fi= \frac{1}{2}\left(\Identity \otimes \Pauli_k + \Pauli_k \otimes \Identity\right) \label{eq:identity_key}
\end{equation}
to transform (\ref{eq:expectation_P-CTCs_5}) into
\begin{align}
	\Expectation_k =\; & \frac{1}{2}\trace\left[\left(\Identity^{1} \otimes \Identity^{2} \otimes \Pauli_k^{3} + \Identity^{1} \otimes \Pauli_k^{2} \otimes \Identity^{3}\right) \UnitaryInteraction^{1,2} \StateCR^{1} {\UnitaryInteraction^{\dagger{1,3}}}\right]. \label{eq:expectation_P-CTCs_6}
\end{align}
Our aim here is to move either term's $\Pauli_k$ operators such that they are on the same mode. Without loss of generality, we choose the third subsystem, which means that we will need to shift the $\Pauli_k$ in the second term onto the third mode. This is easily accomplished by use of \textgate{swap} gates, and following up with a permutation yields the expression
\begin{align}
	\Expectation_k &= \frac{1}{2}\trace\left[\left(\Identity^{1} \otimes \Identity^{2} \otimes \Pauli_k^{3} + \Swap^{2,3} \left\{\Identity^{1} \otimes \Identity^{2} \otimes \Pauli_k^{3}\right\} {\Swap^{\dagger{2,3}}}\right) \UnitaryInteraction^{1,2} \StateCR^{1} {\UnitaryInteraction^{\dagger{1,3}}}\right] \nonumber\\
	&= \trace\left[\Pauli_k^{3} \frac{1}{2}\left(\UnitaryInteraction^{1,2} \StateCR^{1} {\UnitaryInteraction^{\dagger{1,3}}} + {\Swap^{\dagger{2,3}}} \UnitaryInteraction^{1,2} \StateCR^{1} {\UnitaryInteraction^{\dagger{1,3}}}\Swap^{2,3}\right)\right] \nonumber\\
	&= \trace\left[\Pauli_k^{3} \frac{1}{2}\left(\UnitaryInteraction^{1,2} \StateCR^{1} {\UnitaryInteraction^{\dagger{1,3}}} + \UnitaryInteraction^{1,3} \StateCR^{1} {\UnitaryInteraction^{\dagger{1,2}}}\right)\right]. \label{eq:expectation_P-CTCs_7}
\end{align}
The Pauli matrices $\Pauli_k$ on the third subsystem may be moved outside the partial traces on the first and second subsystems,
\begin{align}
	\Expectation_k &= \trace\left\{\Pauli_k \frac{1}{2}\left(\trace_{1,2}\left[\UnitaryInteraction^{1,2} \StateCR^{1} {\UnitaryInteraction^{\dagger{1,3}}}\right] + \trace_{1,2}\left[\UnitaryInteraction^{1,3} \StateCR^{1} {\UnitaryInteraction^{\dagger{1,2}}}\right]\right)\right\} \label{eq:expectation_P-CTCs_8}
\end{align}
which is now in the desired form given by (\ref{eq:expectation_P-CTCs_simplified}).

It is perhaps easier to follow such a calculation when it is presented pictorially in the form of circuit diagrams. Like in the preceding mathematical steps, we begin with the expression (\ref{eq:expectation_P-CTCs}), introduce the third subsystem, use the identity (\ref{eq:identity_key}), and then rearrange particular gates using the \textgate{swap} operation to obtain the desired form. Following convention, we employ a closed loop on a mode to indicate the partial trace of that particular mode.

\begin{align}
	\Expectation_k &= \sum_{n=0}^{1} (-1)^n\trace\left\{\trace_2\left[\UnitaryInteraction\left(\Identity\otimes\ket{n_k}\bra{n_k}\right)\right]\StateCR\,\trace_2\left[\left(\Identity\otimes\ket{n_k}\bra{n_k}\right)\UnitaryInteraction^\dagger\right]\right\}
	\nonumber\\
	&= \sum_{n=0}^{1}(-1)^n \times
	\begin{tikzcd}[row sep={0.25cm}, column sep=0.25cm,transparent]
		\vqw{3} & \qw & \gate[2][0.65cm]{U^\dagger} & \qw & \qw & [-0.25cm] \push{\ \rho\ } & [-0.25cm] \qw & \qw & \gate[2][0.65cm]{U} & \qw & \qw \vqw{3} \\[-0.25cm]
		& \vqw{1} & \qw & \gate[1]{\ket{n_k}\bra{n_k}} & \qw \vqw{1} & & \vqw{1} & \gate[1]{\ket{n_k}\bra{n_k}} & \qw & \qw \vqw{1} & \\
		& & \qw & \qw & \qw & & & \qw & \qw & \qw & \\
		& \qw & \qw & \qw & \qw & \qw & \qw & \qw & \qw & \qw & \qw
	\end{tikzcd}
	\nonumber\\
	&= \sum_{n=0}^{1}(-1)^n \times
	\begin{tikzcd}[row sep={0.25cm}, column sep=0.25cm,transparent]
		\vqw{5} & \qw & \qw & \gate[3,label style={yshift=0.45cm}][1.1cm]{U^{\dagger{1,3}}} & \qw & [-0cm] \push{\ \rho\ } & [-0.25cm] \qw & \qw & \gate[2][1.1cm]{U^{1,2}} & \qw & \qw & \qw & \qw \vqw{5} \\[-0.5cm]
		& \vqw{3} & \qw & \linethrough & \qw & \qw & \qw & \qw & \qw & \gate[1]{\ket{n_k}\bra{n_k}} & \qw & \qw \vqw{3} & \\[-0.2cm]
		& & \vqw{1} & \qw & \qw & \qw & \qw & \qw & \qw & \gate[1]{\ket{n_k}\bra{n_k}} & \qw \vqw{1} & \\
		& & & \qw & \qw & \qw & \qw & \qw & \qw & \qw & \qw & & \\
		& & \qw & \qw & \qw & \qw & \qw & \qw & \qw & \qw & \qw & \qw \\
		& \qw & \qw & \qw & \qw & \qw & \qw & \qw & \qw & \qw & \qw & \qw & \qw
	\end{tikzcd}
	\nonumber\\
	&= \frac{1}{2} \times
	\begin{tikzcd}[row sep={0.25cm}, column sep=0.25cm,transparent]
		\vqw{5} & \qw & \qw & \gate[3,label style={yshift=0.35cm}][1.1cm]{U^{\dagger{1,3}}} & \qw & [-0cm] \push{\ \rho\ } & [-0.25cm] \qw & \qw & \gate[2][1.1cm]{U^{1,2}} & \qw & \qw & \qw & \qw \vqw{5} \\[-0.5cm]
		& \vqw{3} & \qw & \linethrough & \qw & \qw & \qw & \qw & \qw & \qw & \qw & \qw \vqw{3} & \\[-0.5cm]
		& & \vqw{1} & \qw & \qw & \qw & \qw & \qw & \qw & \gate[1]{\sigma_k} & \qw \vqw{1} & \\
		& & & \qw & \qw & \qw & \qw & \qw & \qw & \qw & \qw & & \\
		& & \qw & \qw & \qw & \qw & \qw & \qw & \qw & \qw & \qw & \qw \\
		& \qw & \qw & \qw & \qw & \qw & \qw & \qw & \qw & \qw & \qw & \qw & \qw
	\end{tikzcd}
	+ \frac{1}{2} \times
	\begin{tikzcd}[row sep={0.25cm}, column sep=0.25cm,transparent]
		\vqw{5} & \qw & \qw & \gate[3,label style={yshift=0.35cm}][1.1cm]{U^{\dagger{1,3}}} & \qw & [-0cm] \push{\ \rho\ } & [-0.25cm] \qw & \qw & \gate[2][1.1cm]{U^{1,2}} & \qw & \qw & \qw & \qw \vqw{5} \\[-0.5cm]
		& \vqw{3} & \qw & \linethrough & \qw & \qw & \qw & \qw & \qw & \gate[1]{\sigma_k} & \qw & \qw \vqw{3} & \\[-0.5cm]
		& & \vqw{1} & \qw & \qw & \qw & \qw & \qw & \qw & \qw & \qw \vqw{1} & \\
		& & & \qw & \qw & \qw & \qw & \qw & \qw & \qw & \qw & & \\
		& & \qw & \qw & \qw & \qw & \qw & \qw & \qw & \qw & \qw & \qw \\
		& \qw & \qw & \qw & \qw & \qw & \qw & \qw & \qw & \qw & \qw & \qw & \qw
	\end{tikzcd}
	\nonumber\\
	&= \frac{1}{2} \times
	\begin{tikzcd}[row sep={0.25cm}, column sep=0.25cm,transparent]
		\vqw{5} & \qw & \qw & \gate[3,label style={yshift=0.35cm}][1.1cm]{U^{\dagger{1,3}}} & \qw & [-0cm] \push{\ \rho\ } & [-0.25cm] \qw & \qw & \gate[2][1.1cm]{U^{1,2}} & \qw & \qw & \qw & \qw \vqw{5} \\[-0.5cm]
		& \vqw{3} & \qw & \linethrough & \qw & \qw & \qw & \qw & \qw & \qw & \qw & \qw \vqw{3} & \\[-0.5cm]
		& & \vqw{1} & \qw & \qw & \qw & \qw & \qw & \qw & \gate[1]{\sigma_k} & \qw \vqw{1} & \\
		& & & \qw & \qw & \qw & \qw & \qw & \qw & \qw & \qw & & \\
		& & \qw & \qw & \qw & \qw & \qw & \qw & \qw & \qw & \qw & \qw \\
		& \qw & \qw & \qw & \qw & \qw & \qw & \qw & \qw & \qw & \qw & \qw & \qw
	\end{tikzcd}
	+ \frac{1}{2} \times
	\begin{tikzcd}[row sep={0.25cm}, column sep=0.25cm,transparent]
		\vqw{5} & \qw & \qw & \qw & \gate[3,label style={yshift=0.35cm}][1.1cm]{U^{\dagger{1,3}}} & \qw & [-0cm] \push{\ \rho\ } & [-0.25cm] \qw & \qw & \gate[2][1.1cm]{U^{1,2}} & \qw & \qw & \qw & \qw & \qw \vqw{5} \\[-0.5cm]
		& \vqw{3} & \qw & \gate[swap,style={fill=black!0,draw=black!0,line width=0pt}]{} & \linethrough & \qw & \qw & \qw & \qw & \qw & \gate[swap,style={fill=black!0,draw=black!0,line width=0pt}]{} & \qw & \qw & \qw \vqw{3} & \\[-0.5cm]
		& & \vqw{1} & \qw & \qw & \qw & \qw & \qw & \qw & \qw & \qw & \gate[1]{\sigma_k} & \qw \vqw{1} & \\
		& & & \qw & \qw & \qw & \qw & \qw & \qw & \qw & \qw & \qw & \qw & & \\
		& & \qw & \qw & \qw & \qw & \qw & \qw & \qw & \qw & \qw & \qw & \qw & \qw \\
		& \qw & \qw & \qw & \qw & \qw & \qw & \qw & \qw & \qw & \qw & \qw & \qw & \qw & \qw
	\end{tikzcd}
	\nonumber\\
	&= \frac{1}{2} \times
	\begin{tikzcd}[row sep={0.25cm}, column sep=0.25cm,transparent]
		\vqw{5} & \qw & \qw & \gate[3,label style={yshift=0.35cm}][1.1cm]{U^{\dagger{1,3}}} & \qw & [-0cm] \push{\ \rho\ } & [-0.25cm] \qw & \qw & \gate[2][1.1cm]{U^{1,2}} & \qw & \qw & \qw & \qw \vqw{5} \\[-0.5cm]
		& \vqw{3} & \qw & \linethrough & \qw & \qw & \qw & \qw & \qw & \qw & \qw & \qw \vqw{3} & \\[-0.5cm]
		& & \vqw{1} & \qw & \qw & \qw & \qw & \qw & \qw & \gate[1]{\sigma_k} & \qw \vqw{1} & \\
		& & & \qw & \qw & \qw & \qw & \qw & \qw & \qw & \qw & & \\
		& & \qw & \qw & \qw & \qw & \qw & \qw & \qw & \qw & \qw & \qw \\
		& \qw & \qw & \qw & \qw & \qw & \qw & \qw & \qw & \qw & \qw & \qw & \qw
	\end{tikzcd}
	+ \frac{1}{2} \times
	\begin{tikzcd}[row sep={0.25cm}, column sep=0.25cm,transparent]
		\vqw{5} & \qw & \qw & \gate[2][1.1cm]{U^{\dagger{1,2}}} & \qw & [-0cm] \push{\ \rho\ } & [-0.25cm] \qw & \qw & \gate[3,label style={yshift=0.35cm}][1.1cm]{U^{1,3}} & \qw & \qw & \qw & \qw \vqw{5} \\[-0.5cm]
		& \vqw{3} & \qw & \qw & \qw & \qw & \qw & \qw & \linethrough & \qw & \qw & \qw \vqw{3} & \\[-0.3cm]
		& & \vqw{1} & \qw & \qw & \qw & \qw & \qw & \qw & \gate[1]{\sigma_k} & \qw \vqw{1} & \\
		& & & \qw & \qw & \qw & \qw & \qw & \qw & \qw & \qw & & \\
		& & \qw & \qw & \qw & \qw & \qw & \qw & \qw & \qw & \qw & \qw \\
		& \qw & \qw & \qw & \qw & \qw & \qw & \qw & \qw & \qw & \qw & \qw & \qw
	\end{tikzcd}
	\nonumber\\
	&= \trace\left\{\Pauli_k \frac{1}{2}\left(\trace_{1,2}\left[\UnitaryInteraction^{1,2} \StateCR^{1} {\UnitaryInteraction^{\dagger{1,3}}}\right] + \trace_{1,2}\left[\UnitaryInteraction^{1,3} \StateCR^{1} {\UnitaryInteraction^{\dagger{1,2}}}\right]\right)\right\}.
\end{align}

\section{\uppercase{Equivalence of the na\"{i}ve and tomographical states}}\label{app:equivalency}

As the set of all Pauli matrix (with identity) tensor product pairs $\left\{\Pauli_\mu \otimes \Pauli_\nu\right\}_{\mu,\nu=0}^{3}$ forms a complete basis for the space of complex $4\times4$ matrices, then any unitary matrix $\UnitaryInteraction$ can in general be expanded as
\begin{equation}
	\UnitaryInteraction = \sum_{\mu,\nu=0}^{3}\UnitaryInteraction_{\mu\nu}\Pauli_\mu \otimes \Pauli_\nu. \label{eq:unitary_expansion}
\end{equation}
Here, the factors $\UnitaryInteraction_{\mu\nu}$ are coefficients which describe the particular form of $\UnitaryInteraction$, and the necessary condition of unitarity
\begin{equation}
	\Identity\otimes\Identity = \UnitaryInteraction^\dagger \UnitaryInteraction
\end{equation}
is fulfilled when
\begin{equation}
	\sum_{\mu,\nu = 0}^{3} \UnitaryInteraction^*_{\mu\nu}\UnitaryInteraction_{\mu\nu} = 1. \label{eq:unitarity_statement}
\end{equation}
Our subsequent calculations can be made considerably easier if we first recast (\ref{eq:unitary_expansion}) into an equivalent expression,
\begin{equation}
	\UnitaryInteraction = \sum_{\nu=0}^{3}\UnitaryExpansion_{\nu} \otimes \Pauli_\nu, \label{eq:unitary_expansion_simple}
\end{equation}
where
\begin{equation}
	\UnitaryExpansion_{\nu} \equiv \sum_{\mu=0}^{3}\UnitaryInteraction_{\mu\nu} \Pauli_\mu.
\end{equation}
In this form, the statement of unitarity (\ref{eq:unitarity_statement}) is equivalent to
\begin{align}
	\Identity\otimes\Identity &= \UnitaryInteraction^\dagger \UnitaryInteraction \nonumber\\
	&= \sum_{\mu,\nu=0}^{3}\UnitaryExpansion^\dagger_{\mu}\UnitaryExpansion_{\nu} \otimes \Pauli_\mu \Pauli_\nu \nonumber\\
	&= \UnitaryExpansion^\dagger_{0}\UnitaryExpansion_{0} \otimes \Pauli_0 \Pauli_0 + \sum_{\mu=1}^{3}\UnitaryExpansion^\dagger_{\mu}\UnitaryExpansion_{0} \otimes \Pauli_\mu \Pauli_0 + \sum_{\nu=1}^{3}\UnitaryExpansion^\dagger_{0}\UnitaryExpansion_{\nu} \otimes \Pauli_0 \Pauli_\nu + \sum_{\mu,\nu=1}^{3}\UnitaryExpansion^\dagger_{\mu}\UnitaryExpansion_{\nu} \otimes \Pauli_\mu \Pauli_\nu \nonumber\\
	&= \sum_{\mu=0}^{3}\UnitaryExpansion^\dagger_{\mu}\UnitaryExpansion_{\mu} \otimes \Identity + \sum_{k=1}^{3}\left[\biggl(\UnitaryExpansion^\dagger_{k}\UnitaryExpansion_{0} + \UnitaryExpansion^\dagger_{0}\UnitaryExpansion_{k} + \eye \sum_{\mu,\nu=1}^{3}\epsilon_{\mu\nu k}\UnitaryExpansion^\dagger_{\mu}\UnitaryExpansion_{\nu}\biggr) \otimes \Pauli_k \right] \label{eq:unitarity_statement_expansion}
\end{align}
where we used the fact that $\Pauli_0 \equiv \Identity$ along with the identity
\begin{equation}
	\Pauli_\mu \Pauli_\nu = \delta_{\mu\nu}\Identity + \eye\epsilon_{\mu\nu k} \Pauli_k, \quad \mu,\nu \in \left\{1,2,3,4\right\}.
\end{equation}
Here, $\delta_{\mu \nu}$ is the Kronecker delta, and $\epsilon_{\mu\nu k}$ is the Levi-Civita symbol. For (\ref{eq:unitarity_statement_expansion}) to hold true, we necessarily require both of the following statements to hold true:
\begin{subequations}
	\begin{align}
		\sum_{\mu=0}^{3}\UnitaryExpansion^\dagger_{\mu}\UnitaryExpansion_{\mu} &= \Identity, \\
		\UnitaryExpansion^\dagger_{k}\UnitaryExpansion_{0} + \UnitaryExpansion^\dagger_{0}\UnitaryExpansion_{k} + \eye \sum_{\mu,\nu=1}^{3}\epsilon_{\mu\nu k}\UnitaryExpansion^\dagger_{\mu}\UnitaryExpansion_{\nu} &= 0, \quad \forall \, k. \label{eq:unitary_general_off-diagonal}
	\end{align}
\end{subequations}
With foresight, we can turn (\ref{eq:unitary_general_off-diagonal}) into a more useful form by manipulating the summation term. Permuting the two Greek indices in the Levi-Civita symbol, {\ie}, $\epsilon_{\mu\nu k} = -\epsilon_{\nu\mu k}$, and following this with a label switch $\mu \leftrightarrow \nu$, allows us to write
\begin{equation}
	\sum_{\mu,\nu=1}^{3}\epsilon_{\mu\nu k}\UnitaryExpansion^\dagger_{\mu}\UnitaryExpansion_{\nu} = -\sum_{\mu,\nu=1}^{3}\epsilon_{\nu\mu k}\UnitaryExpansion^\dagger_{\mu}\UnitaryExpansion_{\nu} = -\sum_{\mu,\nu=1}^{3}\epsilon_{\mu\nu k}\UnitaryExpansion^\dagger_{\nu}\UnitaryExpansion_{\mu}.
\end{equation}
With this, we can rearrange (\ref{eq:unitary_general_off-diagonal}) to give
\begin{equation}
	\eye \sum_{\mu,\nu=1}^{3}\epsilon_{\mu\nu k}\UnitaryExpansion^\dagger_{\nu}\UnitaryExpansion_{\mu} = \UnitaryExpansion^\dagger_{k}\UnitaryExpansion_{0} + \UnitaryExpansion^\dagger_{0}\UnitaryExpansion_{k}. \label{eq:unitary_general_off-diagonal_rearranged}
\end{equation}

We can now compute the expectation values for both the weak-measurement $\tilde{\StateCV}$ and na\"{i}ve $\StateCV_\mathrm{P}$ states. For the former, we simply use our unitary expansion (\ref{eq:unitary_expansion_simple}) in the expression (\ref{eq:tomography_P-CTC}) to yield
\begin{align}
	\trace[\Pauli_k \tilde{\StateCV}] &= \frac{1}{2}\trace\left\{ \Pauli_k  \left(\trace_{1,2}\left[\UnitaryInteraction^{1,2} \StateCR^{1} {\UnitaryInteraction^{\dagger{1,3}}}\right] + \trace_{1,2}\left[\UnitaryInteraction^{1,3} \StateCR^{1} {\UnitaryInteraction^{\dagger{1,2}}}\right]\right) \right\}\nonumber\\
	&= \frac{1}{2}\sum_{\mu,\nu=0}^{3} \trace\left[\left(\UnitaryExpansion_\mu^{1} \otimes \Pauli_\mu^{2} \otimes \Identity^{3}\right)\StateCR^{1}\left({\UnitaryExpansion_\nu^{\dagger{1}}} \otimes \Identity^{2} \otimes \Pauli_\nu^{3}\right)\Pauli_k^{3}\right] + \frac{1}{2}\sum_{\mu,\nu=0}^{3} \trace\left[\left(\UnitaryExpansion_\mu^{1} \otimes \Identity^{2} \otimes \Pauli_\mu^{3}\right)\StateCR^{1}\left({\UnitaryExpansion_\nu^{\dagger{1}}} \otimes \Pauli_\nu^{2} \otimes \Identity^{2}\right)\Pauli_k^{3}\right] \nonumber\\
	&= \frac{1}{2}\sum_{\mu,\nu=0}^{3} \trace\left[\UnitaryExpansion_\mu \StateCR \UnitaryExpansion_\nu^\dagger\right] \trace\left[\Pauli_\mu\right] \trace\left[\Pauli_\nu \Pauli_k\right] + \frac{1}{2}\sum_{\mu,\nu=0}^{3} \trace\left[\UnitaryExpansion_\mu \StateCR \UnitaryExpansion_\nu^\dagger\right] \trace\left[\Pauli_\nu\right] \trace\left[\Pauli_\mu \Pauli_k\right] \nonumber\\
	&= 2\sum_{\mu,\nu=0}^{3} \trace\left[\UnitaryExpansion_\mu \StateCR \UnitaryExpansion_\nu^\dagger\right] \delta_{0 \mu} \delta_{\nu k} + 2\sum_{\mu,\nu=0}^{3} \trace\left[\UnitaryExpansion_\mu \StateCR \UnitaryExpansion_\nu^\dagger\right] \delta_{0 \nu} \delta_{\mu k} \nonumber\\
	&= 2\;\trace\left[\bigl(\UnitaryExpansion_k^\dagger \UnitaryExpansion_0 + \UnitaryExpansion_0^\dagger \UnitaryExpansion_k\bigr)\StateCR\right], \label{eq:expectation_tomography}
\end{align}
where we used the cyclic property of the trace, and the identity
\begin{equation}
	\trace\left[\Pauli_\tau \Pauli_k\right] = 2 \delta_{\tau k}.
\end{equation}
For the candidate P-CTC state $\StateCV_\mathrm{P}$ given by (\ref{eq:P-CTCs_CV}), one can use the identity
\begin{equation}
	\trace[\Pauli_n \Pauli_m \Pauli_k] = 2\eye\epsilon_{nmk}
\end{equation}
and the result (\ref{eq:unitary_general_off-diagonal_rearranged}) to similarly compute the expectation values
\begin{align}
	\trace[\Pauli_k \StateCV_\mathrm{P}] &= \frac{1}{2}\trace\left\{\Pauli_k \trace_{1}\left[\UnitaryInteraction^{1,2} \StateCR^{1} {\UnitaryInteraction^{\dagger{1,2}}}\right]\right\} \nonumber\\
	&= \frac{1}{2}\sum_{\mu,\nu=0}^{3} \trace\left[\left(\UnitaryExpansion_\mu^{1} \otimes \Pauli_\mu^{2}\right)\StateCR^{1}\left({\UnitaryExpansion_\nu^{\dagger{1}}} \otimes \Pauli_\nu^{2}\right)\Pauli_k^{2}\right] \nonumber\\
	&= \frac{1}{2}\sum_{\mu,\nu=0}^{3} \trace\left[\UnitaryExpansion_\mu \StateCR \UnitaryExpansion_\nu^\dagger\right] \trace\left[\Pauli_\mu \Pauli_\nu \Pauli_k\right] \nonumber\\
	&= \frac{1}{2} \trace\bigl[\UnitaryExpansion_0 \StateCR \UnitaryExpansion_0^\dagger\bigr] \trace\left[\Pauli_0 \Pauli_0 \Pauli_k\right] + \frac{1}{2}\sum_{\nu=1}^{3} \trace\left[\UnitaryExpansion_0 \StateCR \UnitaryExpansion_\nu^\dagger\right] \trace\left[\Pauli_0 \Pauli_\nu \Pauli_k\right] \nonumber\\
	&\quad + \frac{1}{2}\sum_{\mu=1}^{3} \trace\bigl[\UnitaryExpansion_\mu \StateCR \UnitaryExpansion_0^\dagger\bigr] \trace\left[\Pauli_\mu \Pauli_0 \Pauli_k\right] + \frac{1}{2}\sum_{\mu,\nu=1}^{3} \trace\left[\UnitaryExpansion_\mu \StateCR \UnitaryExpansion_\nu^\dagger\right] \trace\left[\Pauli_\mu \Pauli_\nu \Pauli_k\right] \nonumber\\
	&= 0 + \trace\bigl[\UnitaryExpansion_0 \StateCR \UnitaryExpansion_k^\dagger\bigr] + \trace\bigl[\UnitaryExpansion_k \StateCR \UnitaryExpansion_0^\dagger\bigr] + \eye\sum_{\mu,\nu=1}^{3} \epsilon_{\mu\nu k}\trace\left[\UnitaryExpansion_\mu \StateCR \UnitaryExpansion_\nu^\dagger\right] \nonumber\\
	&= \trace\left[\biggl(\UnitaryExpansion_k^\dagger \UnitaryExpansion_0 + \UnitaryExpansion_0^\dagger \UnitaryExpansion_k + \eye\sum_{\mu,\nu=1}^{3} \epsilon_{\mu\nu k}\UnitaryExpansion_\nu^\dagger \UnitaryExpansion_\mu \biggr)\StateCR\right] \nonumber\\
	&= 2\;\trace\left[\bigl(\UnitaryExpansion_k^\dagger \UnitaryExpansion_0 + \UnitaryExpansion_0^\dagger \UnitaryExpansion_k\bigr)\StateCR\right]. \label{eq:expectation_naive}
\end{align}

\end{document}